\documentclass[twocolumn]{aastex631}
\usepackage{amsmath}
\usepackage{bm}
\usepackage{graphicx}
\usepackage{CJK}

\DeclareFontFamily{U}{matha}{\hyphenchar\font45}
\DeclareFontShape{U}{matha}{m}{n}{
      <5> <6> <7> <8> <9> <10> gen * matha
      <10.95> matha10 <12> <14.4> <17.28> <20.74> <24.88> matha12
      }{}
\DeclareSymbolFont{matha}{U}{matha}{m}{n}

\DeclareMathSymbol{\Lt}{3}{matha}{"CE}
\DeclareMathSymbol{\Gt}{3}{matha}{"CF}

\begin{document}
\begin{CJK*}{UTF8}{gbsn}
\title{Radiation-magnetohydrodynamic Simulations of Accretion Flow Formation After a Tidal Disruption Event}

\author{Maria Renee Meza}
\affil{Department of Astronomy, University of Virginia, 530 McCormick Road, Charlottesville, VA 22904, USA}
\email{qqn8hw@virginia.edu}

\author{Xiaoshan Huang (黄小珊)}
\affil{California Institute of Technology, TAPIR, Mail Code 350-17, Pasadena, CA 91125, USA}

\author{Shane W. Davis}
\affil{Department of Astronomy, University of Virginia, 530 McCormick Road, Charlottesville, VA 22904, USA}
\affil{Virginia Institute for Theoretical Astronomy, University of Virginia, Charlottesville, VA 22904, USA}

\author{Yan-Fei Jiang (姜燕飞)}
\affil{Center for Computational Astrophysics, Flatiron Institute, 162 Fifth Avenue, New York, NY 10010, USA}

\begin{abstract}
We perform 3D radiation-magnetohydrodynamic simulations of the evolution of the fallback debris after a tidal disruption event. We focus on studying the effects of magnetic fields on the formation and early evolution of the accretion flow. We find that large magnetic fields can increase the debris stream thickness, moderately reducing the efficiency of the radiative acceleration of outflows during the first self-intersecting collisions. As gas accumulates and the collisions happen instead between the infalling stream and the accretion flow, magnetized and nonmagnetized systems evolve similarly at these early times: radiation-driven outflows dominate early after the initial stream-stream collision and a few days later, the accretion rate exceeds the mass outflow rate. We find that the MRI does not play a significant role in angular momentum transport and dissipation. Nor do we find evidence of a magnetocentrifugal driven outflow. Instead, collisions continue to dissipate kinetic energy into radiation that launches outflows and powers TDE luminosities reaching $L\sim4-6\times10^{44}$ erg s$^{-1}$. Shock-driven outflows and inflows redistribute angular momentum throughout the extent ($\sim50 r_s$) of the forming eccentric disk. Even in the presence of magnetic stresses, the accretion flow remains mostly eccentric with $e\sim0.2-0.3$ for $r\lesssim8r_s$ and $e\sim0.4-0.5$ for $10\lesssim r\,(r_s)\lesssim50$. Lastly, we find a polar angle-dependent density structure compatible with the viewing-angle effect, along with an additional azimuthal angle dependence established by the collisions.
\end{abstract}


\section{Introduction}
\label{s:introduction}
\end{CJK*}
Tidal disruption events (TDEs) are luminous transients observed after a star in a highly eccentric orbit passes too close to a super massive black hole (SMBH) and gets disrupted by the tidal forces. If the star is fully disrupted, a large fraction of the stellar mass will remain bound to the black hole forming a stream of debris. As this debris falls back to the black hole on a nearly ballistic orbit, it will eventually intersect itself due to relativistic apsidal precession. This collision dissipates kinetic energy through shocks and begins the circularization of gas. Early theoretical work predicted the rapid formation of an accretion disk fed at a rate of $\dot{M} \propto t^{-5/3}$, equal to the rate at which the bound stellar debris returns to the pericenter after the disruption \citep{Rees1988,Phinney1989}. 

TDEs have been observed across optical \citep{vanVelzen2011,Arcavi2014,Holoien2019,Nicholl2019,van2021seventeen,Hammerstein2023,Yao2023,Yao2025}, UV \citep{Gezari2006,Gezari2008, Gezari2009, Saxton2017,Blagorodnova2019}, X-ray \citep{Donley2002,Bloom2011,Jonker2020,Eftekhari2024,Guolo2024}, and radio wavelengths \citep{Zauderer2011,Cenko2012,alexander2020radio,goodwin2023radio,Cendes2024,Dykaar2024,somalwar2025vlass}. Several optical/UV detected TDEs have light curves that show a rise in a timescale of weeks and a decline proportional to $t^{-5/3}$ across several months \citep{Gezari2021}. This power-law decline has also been observed for a few X-ray detected TDEs \citep{Halpern2004}. This led to a variety of accretion-powered models to explain the luminous emission from TDEs. \cite{Dai2015} proposed that the deepest disruptions as quantified by the impact parameter $\beta=r_t/r_p$, where $r_t$ and $r_p$ are the tidal and pericenter radius, respectively, would be the ones to emit in soft X-rays. A fraction of TDEs show both optical and  X-ray emission \citep{Kajava2020, Guolo2024, Malyali2024}. To explain this population, reprocessing outflow models have been proposed. In these models, the outflows launched from super-Eddington accretion reprocess X-ray photons into optical/UV emission \citep{Roth2016, Dai2018}. The X-ray emission is then explained by the outflows becoming ionized and, therefore, transparent to X-ray photons and may also involve viewing-angle effects \citep{Metzger2016, Dai2018,Thomsen2022}. In the viewing-angle effect model, optical emission is detected in lines of sight near the midplane of the accretion disk, while X-ray emission is preferentially observed near the poles \citep{Dai2018,Parkinson2024}.  By fitting TDE X-ray spectra with a relativistic accretion disk model, \cite{Mummery2023} showed that their early optical luminosities cannot be explained by reprocessing alone, as their peak X-ray disk luminosities were smaller for some sources. Importantly, these accretion-powered models require the prompt formation of an accretion disk after the return of the most bound debris to the black hole. However, by studying the effects of relativistic precession through Monte Carlo methods, \cite{Guillochon2015} found that accretion may be significantly delayed due to large viscous timescales for $M_{BH}\lesssim10^6\, M_{\odot}$.  Notably, this work found that identifications may therefore be biased toward prompt TDEs, since delayed events will show significantly slower evolving light curves.

Alternatively, the luminous emission of TDEs has been attributed to the accretion flow formation process rather than accretion itself \citep{Piran2015}. Local \citep{Jiang2016, Huang2023} and global \citep{Huang2024} radiation hydrodynamic simulations of the evolution of the fallback debris have shown that stream-stream collisions dissipate kinetic energy into radiation, reaching luminosities of $\sim10^{42}-10^{44}$ ergs s$^{-1}$, consistent with observed bolometric luminosities of TDEs. These powerful shocks launch asymmetric, optically-thick outflows with photospheric temperatures $T\sim10^4\, K$  \citep{Huang2024}, similar to the temperatures measured for optical TDEs. Simulations that follow the disruption of the star and its long-term evolution have also found the launching of outflows, which modulate the accretion rate \citep{Ryu2023, Price2024}. Stream-stream and stream-disk collisions have been shown to be able to power TDEs while the accretion flow formed through this dissipation mechanism remains eccentric \citep{Shiokawa2015,Ryu2023, Huang2024, Steinberg2024}, unlike the circularized Keplerian disk often assumed in some accretion-powered models. 

Previous simulation studies have mostly focused on the (radiation) hydrodynamic evolution of TDEs. However, the stellar magnetic field will thread the debris after the disruption; consequently, magnetic forces will be involved in the evolution of the system. In particular, we can expect that magnetic pressure might end confinement by self-gravity, widening the stream debris \citep{Bonnerot2017_2}. The expanded debris structure may then affect the dissipation during the stream-stream and stream-disk collisions. Furthermore, we can expect magnetic stresses to participate in the subsequent accretion flow. In standard accretion disk theory, magnetic fields play a central role through the magnetorotational instability (MRI), which drives accretion through turbulent stress \citep{Balbus1991, Balbus1999}. Notably,  magnetized eccentric disks show steeper eccentricity gradients than their nonmagnetized counterparts, which may affect the radiative efficiency expected from accretion \citep{Chan2022}. 

The magnetohydrodynamical simulations of TDEs by \cite{Guillochon2017}, exploring two magnetic field strengths and geometries, showed that after disruption, the magnetic field lines tend to align with the direction of elongation. Furthermore, they found that since the magnetic pressure declines slower than the thermal pressure, the former may dominate over the latter in a timescale comparable to that in which hydrogen may recombine, depending on the magnetic field strength. In agreement, \cite{Bonnerot2017_2} found that the field lines align with the stretching direction, except in the case where the stellar magnetic field was strictly perpendicular to the trajectory of the star. They found that the magnetic field becomes dynamically important a few tens of hours after disruption if the star had a large magnetic field $B_*\geq1$ MG. There is still uncertainty about the magnetic fields in stellar interiors \citep{Fisher2000,Brun2004,Vasil2024}. For the Sun, magnetic fields strengths of $\sim3\times10^4-10^5$ G have been estimated at the base of the convective zone \citep{Fisher2000}. For the near-surface shear layer, toroidal field strengths of $\sim380-1400$ G have been inferred from helioseismology \citep{Baldner2009}. The mean photospheric field of the Sun has been measured to be relatively weak $\sim7.7$ G \citep{Kotov2008}. 

Other studies have followed the accretion flow formation of TDEs using magnetohydrodynamic simulations. \cite{Sadowski2016} modeled the close disruption of a red dwarf by a $M_{BH}=10^5M_{\odot}$ black hole using a smoothed particle hydrodynamics code and followed the subsequent evolution using the magnetohydrodynamical code \texttt{KORAL}. They found that the Reynolds stress is an order of magnitude larger than the Maxwell stress in their formed eccentric accretion disk. \cite{Curd2021}, also using \texttt{KORAL}, focused on the accretion flow formation stage, after close disruptions, by injecting the stream into the simulation grid. For their weakly magnetized run, they found that the ratio of the magnetic pressure to the sum of gas and radiation pressure remained small after circularization, leading to a similar conclusion as \cite{Sadowski2016} that the magnetic field did not significantly impact the dynamics of the system.

In this study, we focus on further investigating the role of magnetic fields in the accretion and outflow dynamics following a TDE. In the two previous studies, weakly magnetized systems have been considered. Resolving the effects of weak magnetic fields requires exceptionally high resolution, which proves to be challenging given the large range of length scales relevant in TDE systems. In this work, we focus on resolving the potential magnetic effects on the dynamics of TDE accretion flow formation by considering a larger magnetic field. We do this by modeling the debris stream on its trajectory back to the SMBH through radiation-magnetohydrodynamic (RMHD) simulations. This paper is organized as follows. In Section \ref{s:methods}, we describe the setup and initial conditions of our three simulations. In Section \ref{s:results}, we report our results. In particular, in Sections \ref{ss:accretion} and \ref{ss:outflows} we describe how magnetic fields impact accretion and outflows, respectively. Next, we measure the luminosity and the radiative and kinetic efficiencies in Section \ref{ss:radiation}. In Section \ref{ss:angmom}, we analyze the redistribution of angular momentum throughout the accretion flow. Lastly, we examine the structure of the forming accretion disk in Section \ref{ss:structure}. We discuss our results and compare them with previous works in Section \ref{s:discussion}, and conclude in Section \ref{s:conclusions}.

\section{Methods}
\label{s:methods}

We set up our simulations using the \texttt{Athena++} \citep{Stone2020} code. \texttt{Athena++} is a finite-volume, flux-conservative radiation magnetohydrodynamics code that solves the radiative transfer equation directly. Similar to \cite{Jiang2014}, we integrate the transport operator explicitly, but solve the source term implicitly via operator splitting. The code evolves the magnetic fields through the constrained transport method \citep{Evans1988}, which ensures $\nabla\cdot \bm{B}=0$.  We use the HLLD Riemann solver with second-order spatial reconstruction. The equations solved by the code are \citep{Jiang2021},
\begin{equation}
    \frac{\partial \rho}{\partial t}+\nabla\cdot(\rho\bm{v})=0,
\end{equation}
\begin{equation}
    \frac{\partial\rho\bm{v}}{\partial t}+\nabla\cdot(\rho\bm{vv}+P\tilde{I}-\bm{BB})=\bm{G}+\rho\bm{a_g},
    \label{eq:mom}
\end{equation}
\begin{equation}
    \frac{\partial E}{\partial t}+\nabla\cdot\big[(E+P\tilde{I})\bm{v}-\bm{B}(\bm{B}\cdot\bm{v})\big]=cG_0+ \rho\bm{a_{g}}\cdot\bm{v},
\end{equation}
\begin{equation}
    \frac{\partial\bm{B}}{\partial t}-\nabla\times(\bm{v}\times\bm{B})=0,
\end{equation}
\begin{equation}
    \frac{\partial I}{\partial t}+c\bm{n}\cdot\nabla I=cS_I,
\end{equation}
\begin{equation}
    \begin{aligned}
    S_I\equiv&\Gamma^{-3}\Bigg[\rho(\kappa_s+\kappa_a)(J_0-I_0)\\
    &+\rho\kappa_P\bigg(\frac{c aT^4}{4\pi}-J_0\bigg)\Bigg],
    \end{aligned}
\end{equation}
\begin{equation}
    \bm{G}\equiv -\frac{1}{c}\int \bm{n}S_I d\Omega,
\end{equation}
\begin{equation}
    G_0\equiv -\frac{1}{c}\int S_Id\Omega,
\end{equation}

which are the mass, momentum, and energy conservation equations, the induction equation, and the radiative transfer equation. The variables are the density $\rho$, velocity $\bm{v}$, the pressure tensor $P\tilde{I}=(P_{gas}+B^2/2)\tilde{I}$ where $\tilde{I}$ is the identity matrix, the magnetic field $\bm{B}$, the sum of the kinetic, thermal, and magnetic energy densities $E$, the intensity $I$, and the unit direction vector of the radiation field $\bm{n}$. $G_0$ and $\bm{G}$ are the time and spatial components, respectively, of the radiation four-force, and $\Gamma$ is the Lorentz factor. $\bm{a_g}$ is the acceleration given by the \cite{Tejeda2013} generalized Newtonian potential $\Phi_{TR}$. For matter-radiation interactions, we consider electron scattering opacity $\kappa_{es}=0.34$ cm$^{2}$g$^{-1}$ for a solar composition, and use the Rosseland and Planck mean opacity tables by \cite{Zhu2021}, which include dust, molecular, and atomic opacities for the range of densities and temperatures of interest. These set the Planck mean $\kappa_P$ and absorption opacities $\kappa_{a}$. For temperatures exceeding the maximum tabulated temperature, we scale the Planck mean opacity using a Kramer's law $\propto T^{-3.5}$. The radiative transfer equation is solved across $80$ angles.

Our setup is similar to that in \cite{Huang2024}: we model the fallback stream of debris by injecting gas at a fixed rate into a spherical polar grid ($r,\theta,\varphi$) and follow its evolution as it returns to the black hole. Note that in this study, the injection happens at the radial boundary instead of the active grid. To define the initial conditions, we start by numerically computing a ballistic orbit (overplotted in green in Figure \ref{fig:3}) around a $M_{BH}=3\times10^6M_{\odot}$ black hole using the generalized Newtonian potential by \cite{Tejeda2013}, which can exactly reproduce the relativistic apsidal precession and the location of the innermost stable circular orbit (ISCO), among other features, of Schwarzschild geodesics. We also use this potential as the explicit gravitational source term in our RMHD simulations. We consider a Sun-like star was disrupted with an impact parameter $\beta=r_t/r_p=1.73$, where $r_t$ and $r_p$ are the tidal and pericenter radius, respectively, roughly corresponding to a full disruption \citep{Guillochon2013}. The eccentricity of the most bound gas can be estimated as in \cite{Dai2015}
\begin{equation}
    e=1-2\beta^{-1}\Big(\frac{M_*}{M_{BH}}\Big)^{1/3},
    \label{eq:1}
\end{equation}
where $M_*$ is the original mass of the disrupted star and $M_{BH}$ is the black hole mass. For our parameters, this yields $e=0.99$. From this orbit, we select a radius that coincides with our grid boundary $r\sim400r_s$, where $r_s$ is the Schwarzschild radius; this sets the position $\bm{r_{inj}}=(402 r_s,\pi/2, 0.22)$ and velocity $\bm{v_{inj}}=(-0.04c, 0, 0.006c)$ of the injected stream. We inject the stream at the radial boundary, where the injection zone consists of $2\times2\times3$ $(r,\theta,\varphi)$ cells which are selected based on a proximity criterion from $\bm{r_{inj}}$. The stream of solar composition ($\mu=0.6$) and temperature $T=5\times10^4$ K is injected at a constant rate of $\dot{M}_{fb}=10\dot{M}_{Edd}$, where $\dot{M}_{Edd}=40\pi GM_{bh}/k_{es}c$ assuming a radiative efficiency $\eta=0.1$. We choose $\dot{M}_{fb}=10\dot{M}_{Edd}$ as a representative fallback rate based on the fallback rate curve for our parameters generated using the STARS library \citep{Law-Smith2020, STARS2020}. We model the density profile of the stream as $\rho=\rho_0\exp{\big(-d^2/r_0^2\big)}$, where $d$ is the distance from the coordinates of the injection cell to $\bm{r_{inj}}$. The parameter $\rho_0$ is set to ensure the total mass injection rate $\dot{M}_{fb}$ through the injection zone and $r_0=3.7\,r_s$ is of the order of the injection zone size. 

Our base grid has a resolution of $64\times32\times64$ cells, but using five levels of adaptive mesh refinement \citep{Stone2020} based on density, density gradient, and distance from the midplane criteria, we reach an effective resolution of $1024\times512\times1024$. We chose this level of refinement based on the resolution study by \cite{Huang2024} concerning the stream width during pericenter passage. Our grid spans $2.7\leq r (r_s)\leq400$, $0\leq\theta\leq\pi$, and $0\leq\varphi\leq2\pi$. Except for the injection zone, the outer radial boundary conditions are set to outflow only; the inner radial boundary conditions are set to inflow only; the $\theta$-boundaries use polar boundary conditions, which copy the cell variables to the cell $180$ degrees from it across the pole; and the azimuthal boundaries are periodic. The active grid is initialized to floor values.

We produce three simulations that only differ by the magnetic field of the injected stream. The stream in the fiducial RMHD simulation, is threaded by a $\sim2600$ G average magnetic field set by the vector potential $A_{\theta}=B_0r_0\exp(-d^2/r_0^2)$, following the density structure of the stream. Here $B_0$ is a parameter set to enforce a plasma beta, the ratio of the thermal pressure to the magnetic pressure, $\beta_M=P_{gas}/P_{mag}\sim0.05$. The second RMHD simulation has the same magnetic field strength and average $\beta_M$, but its structure is set by the vector potential $A_{\varphi}=B_0r_0\exp(-d^2/r_0^2)$. The only difference between these two runs then is the magnetic field geometry; the injected magnetic field in the fiducial run is toroidal, while in the second RMHD simulation the injected field is poloidal. The first field geometry is motivated by the findings of \cite{Guillochon2017} and \cite{Bonnerot2017_2}, who found that as long as the original stellar magnetic field has a component parallel to the orbital plane, the stream magnetic field lines will tend to align with the direction of elongation. The latter geometry is explored for comparison, since the magnetorotational instability is known to have larger growth rates in the presence of net vertical fields \citep{Balbus1998}.

After disruption, as the stream elongates and expands, the magnetic field strength is expected to decrease according to magnetic flux conservation in ideal magnetohydrodynamics. In their simulations, \cite{Bonnerot2017_2} found that the field strength had decreased from $B_*=1$ G to $B=0.1$ G $20$ hours after disruption. However, properly resolving the MRI wavelength of a very weak field is not possible in our setup given the general, large length scales of the problem. Therefore, the low $\beta_M$ was chosen to ensure that we resolve the magnetorotational instability, known to drive mass accretion in standard accretion disks \citep{Balbus2003}, since a larger magnetic field in lieu of even finer resolution may increase the quality factors \citep{Hawley2013}. Note that the instability may still develop even for these larger magnetic field strengths \citep{Balbus1998,Kim2000}. In the third simulation, the stream does not have a magnetic field to facilitate the isolation of the magnetic field effects on TDE evolution. Throughout this work, we will refer to the RMHD simulations as MHD-T (for a toroidal field), MHD-P (for a poloidal field), and HD (for no magnetic field), respectively. We ran the simulations up to $7$ days after the initial stream collision. By this time, a total mass of $0.0195M_{\odot}$ has been injected through the grid boundary. The length unit is the Schwarzschild radius $r_s=2GM_{BH}/c^2=8.86\times10^{11}$ cm.

\section{Results}
\label{s:results}

As the stream first falls toward the black hole, its dynamics is determined mainly by the orbital motion. However, for the MHD runs, the additional magnetic pressure support resists tidal compression, making the infalling stream considerably thicker than the stream in HD, which is supported by radiation pressure. As the stream passes the pericenter, it is compressed and is deflected due to relativistic apsidal precession. In all runs, this compression raises the temperature of the stream, and its structure after pericenter passage is then set by the tidal gravity and radiation pressure support. As it moves away from the black hole, it re-expands and in its return to its new apocenter, the returning stream collides with the still-infalling stream. This results in a strong shock that dissipates some of the kinetic energy of the stream into radiation. The high radiation pressure promptly accelerates gas into fast outflows. We find more outflowing gas in HD compared to the MHD runs. This is because the stream collision in HD happened between two very dense, thin stream sections. In contrast, in the MHD runs, the collision occurred between the thick and diffuse infalling stream and the thin returning stream. Therefore, the radiation produced during the collision could more efficiently accelerate the gas in the more opaque collision zone of HD. This expansion of the stream by the magnetic force is then a distinguishing feature between the magnetized and nonmagnetized runs. During the collision, some gas is not radiatively accelerated, but rather, as its kinetic energy is dissipated, it becomes more bound to the black hole. These shocks repeat themselves over the whole duration of the simulation, but they get weaker each time, as the returning stream increasingly interacts with the growing accumulated gas near the black hole instead of the infalling stream. We note that this transition from stream-stream collisions to stream-disk\footnote{Note that we use the term ``stream-disk" collision to denote the interaction of the stream with the accumulated gas near the black hole, despite the lack of a well-defined disk. Throughout this paper, the term ``disk" is also used when referring to the forming accretion flow.} collisions begins approximately after the first collision for the MHD runs, and after the third collision for HD.  The accumulating gas also interacts with the infalling stream as it orbits the black hole. These constant interactions between the stream and the accumulating gas circularize the debris, and we can see a more disk-like structure beginning to form by the end of the runs.

\begin{figure*}
    \centering
    \includegraphics[width=\textwidth]{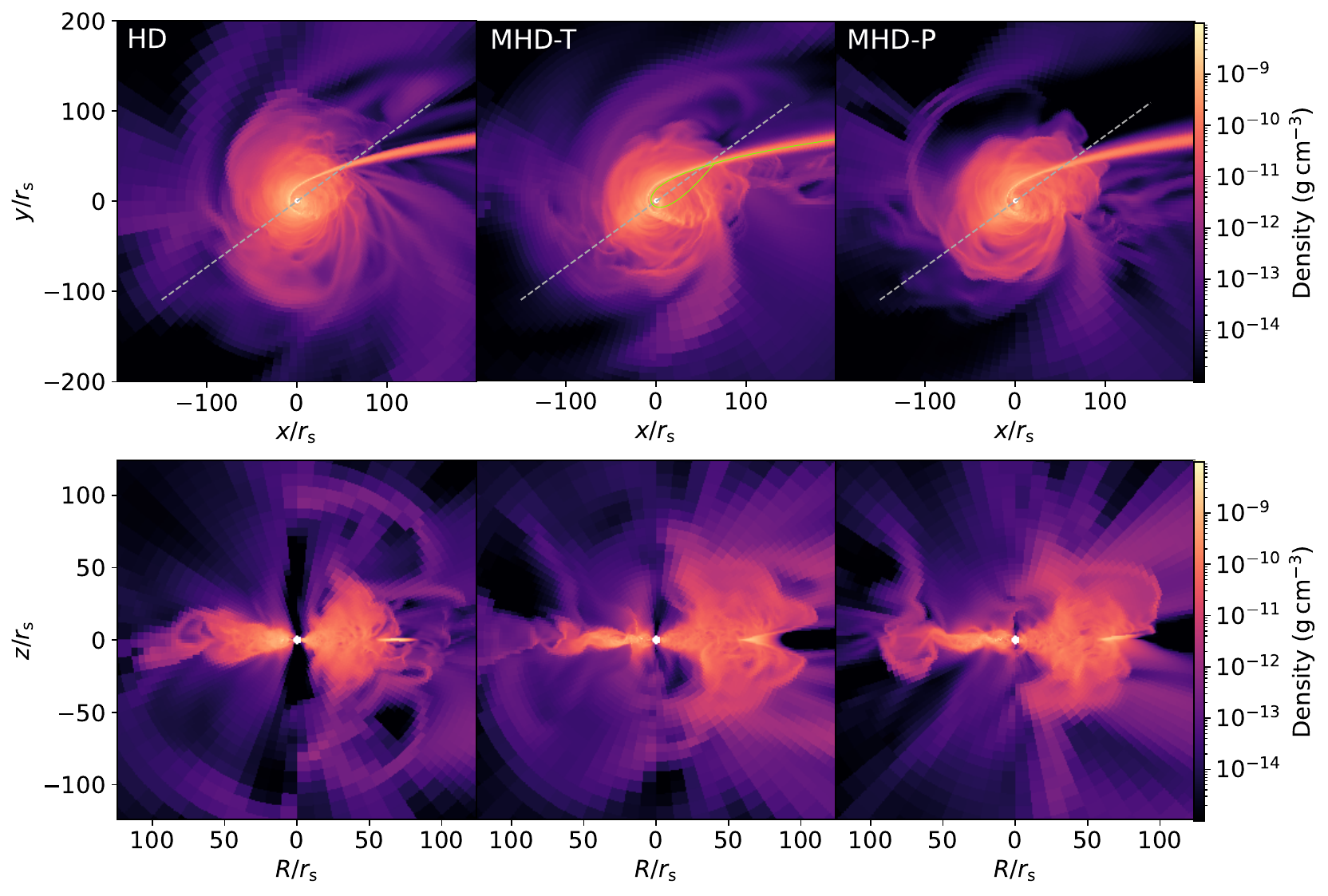}
    \caption{From left to right: Density maps of HD, MHD-T, and MHD-P at the end of the simulations, $t=7$ days. The top row shows the midplane view, and the bottom row shows an azimuthal slice at the collision angle. The location of the azimuthal slice is marked as a grey dashed line on the midplane view panels. The original ballistic orbit is overplotted in green in the top MHD-T panel.}
    \label{fig:3}
\end{figure*}

We show a snapshot of the gas density at the end of our simulations in Figure \ref{fig:3}. In the three runs, an eccentric accretion flow extending mostly out to $\sim50\,r_s$ can be seen from the midplane view, while the azimuthal slice shows asymmetric outflows. Additionally, the MHD runs seem to have more vertically extended structures than HD in the collision zone.

\begin{figure*}
    \centering
    \includegraphics[width=\textwidth]{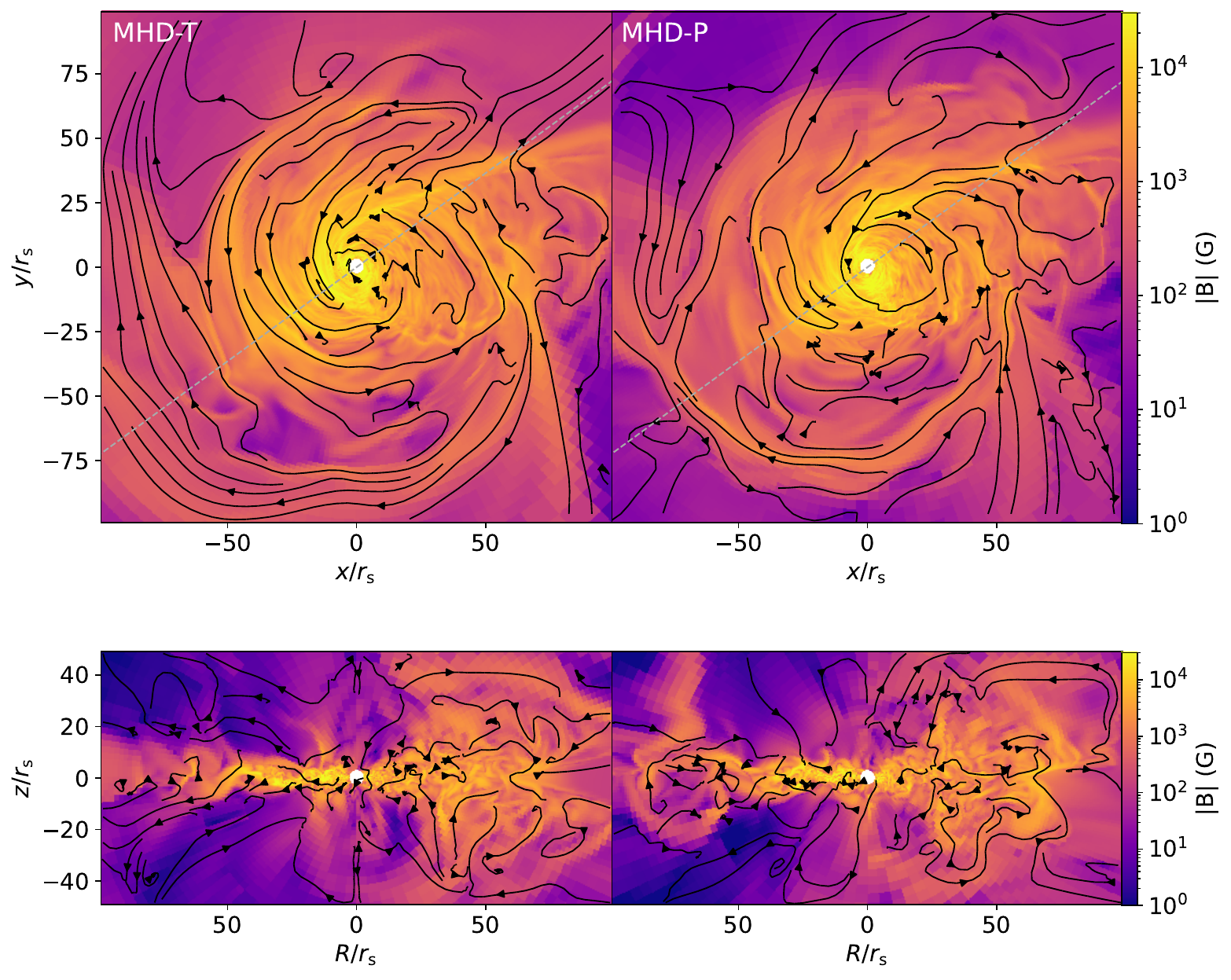}
    \caption{Magnetic field strength at the end of the runs for MHD-T (left) and MHD-P (right). The streamlines show the projected magnetic field lines. The top row shows the midplane view, and the bottom row shows an azimuthal slice at the collision angle.}
    
    \label{fig:bfield}
\end{figure*}

Figure \ref{fig:bfield} shows the magnetic field strength at the end of the MHD runs and the projected magnetic field lines. The magnitude of the magnetic field is enhanced by factors of $\sim2-10$ near the black hole $r\lesssim50r_s$ and close to the midplane, compared to the original field strength of the injected stream. The magnetic field is highly turbulent and shows numerous reversals. This is due to the persistent shocks within the accretion flow and the system being largely radiation-pressure dominated. The average toroidal magnetic energy density throughout the disk is larger than the other components by factors $\sim2-10$ for both runs.

Throughout this paper, the time $t$ is presented as the time since the start of the first stream-stream collision, which happens $\sim2.13$ days after the stream is first injected into the grid. At this point, we set $t=0$.

\subsection{Quality factors}
\label{ss:qfactors}
We compute the quality factors to ensure that we are resolving the magnetorotational instability in the MHD simulations. We calculate both $Q_{\varphi}=2\pi v_{a,\varphi}\big/\Omega r\sin{\theta}\Delta\varphi$ and $Q_{\theta}=2\pi v_{a,\theta}\big/\Omega r\Delta\theta$ where $v_a$ is the Alfven speed and $\Omega$ is the magnitude of the angular frequency at a given radius. We compute the mass-weighted, azimuthal and time averages for the gas within $10$ degrees (0.1745 radians) from the midplane for the last two days of the simulation. For both MHD simulations, we find $\langle\langle Q_\varphi\rangle\rangle\sim18-24$ for $r\lesssim10r_s$ and $\langle\langle Q_\varphi\rangle\rangle\sim18-20$ for $10\lesssim r(r_s)\lesssim100$, while $\langle\langle Q_\theta\rangle\rangle\sim7$ for $r\lesssim10r_s$, increasing with radius up to $\langle\langle Q_\theta\rangle\rangle\sim12$ and $\langle\langle Q_\theta\rangle\rangle\sim18$ for $10\lesssim r\lesssim100r_s$ in MHD-T and MHD-P, respectively. Given these values, we conclude that we are moderately resolving the MRI in our simulations \citep{Hawley2013}. Although we have a relatively strong magnetic field, we verified that the MRI wavelength $\lambda_{MRI}=2\pi v_{a,\theta}/\Omega$ fits within a scale height of the forming accretion flow. 

\subsection{Accretion}
\label{ss:accretion}

One of the questions we are interested in exploring is how the magnetic fields can impact accretion in TDEs. We compute the accretion rate at $r=3r_s$, which corresponds to the ISCO radius of a Schwarzschild black hole, using

\begin{equation}
    \dot{M}=\int_A\rho v_r r^2\sin\theta d\theta d\varphi.
    \label{eq:2}
\end{equation}
This corresponds to the total mass per unit time passing through a spherical shell of radius $r$.

\begin{figure}
    \centering
    \includegraphics[width=0.5\textwidth]{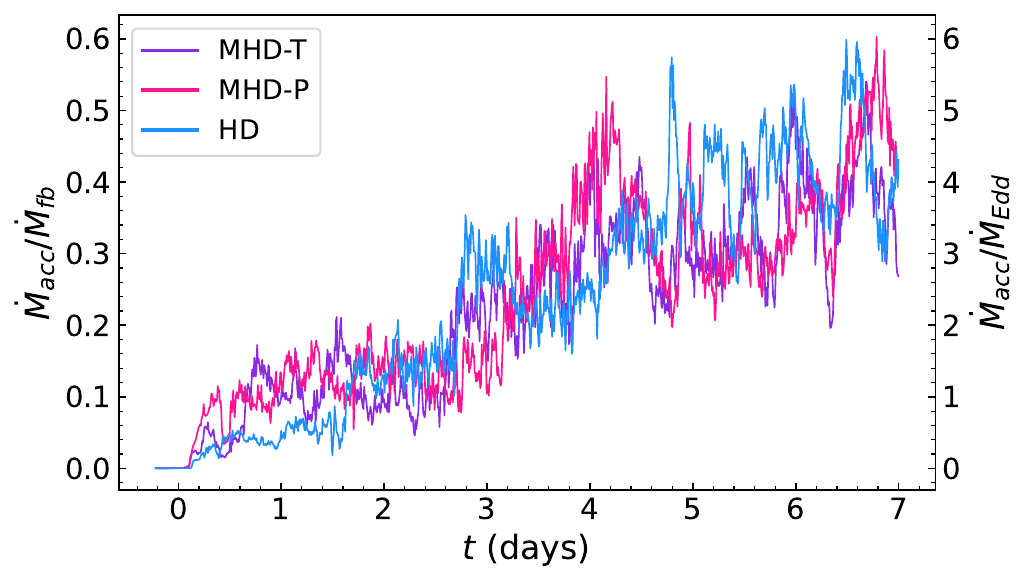}
    \caption{Mass accretion rate measured at $r=3r_s$ as a function of time since the first stream self-intersection.}
    \label{fig:4}
\end{figure}

Figure \ref{fig:4} shows the accretion rate for the three simulations as a function of time. We can see that the accretion rate is initially $2-6$ times higher for the MHD runs than for HD up to $t\sim1.7$ days. In the MHD runs, the initial stream-stream collisions happened between a diffuse and a dense stream section, and the less opaque collision zone made the radiative acceleration of gas less efficient than in HD. Therefore, a larger fraction of shocked gas was promptly accreted onto the black hole in the MHD runs. The mass accretion varies stochastically on very short time scales, because of this, after $t\sim1.7$ days there is not a clear difference between the accretion rate trends among the runs. However, on average, the accretion rates are increasing with time, varying mostly between $\sim5-20\%$ of the fallback rate ($10\dot{M}_{Edd}$) initially and between $\sim30-50\%$ by the end of the runs. This implies super-Eddington accretion $\sim3-5\dot{M}_{Edd}$.

To investigate the processes that dominate the mass accretion in these systems, we computed the Reynolds and Maxwell stresses for the inner $100r_s$. We compute the Reynolds stress following \cite{Jiang2019}, subtracting the mean inflow velocity in the radial and azimuthal directions. A derivation in spherical polar coordinates can be found in Section \ref{ss:angmom}. We average the stresses in time and azimuth for the gas within $10$ degrees from the midplane. The time average is performed for the range $5\leq t\leq7$, which corresponds to the last two days of the simulation. The stresses are normalized by the same averaging of the sum of the radiation and gas pressures. Within these averages, the radiation pressure is always $\sim2$ and $\sim3-4$ orders of magnitude larger than the magnetic and gas pressures, respectively, across the runs. 

\begin{figure*}
    \centering
    \includegraphics[width=\textwidth]{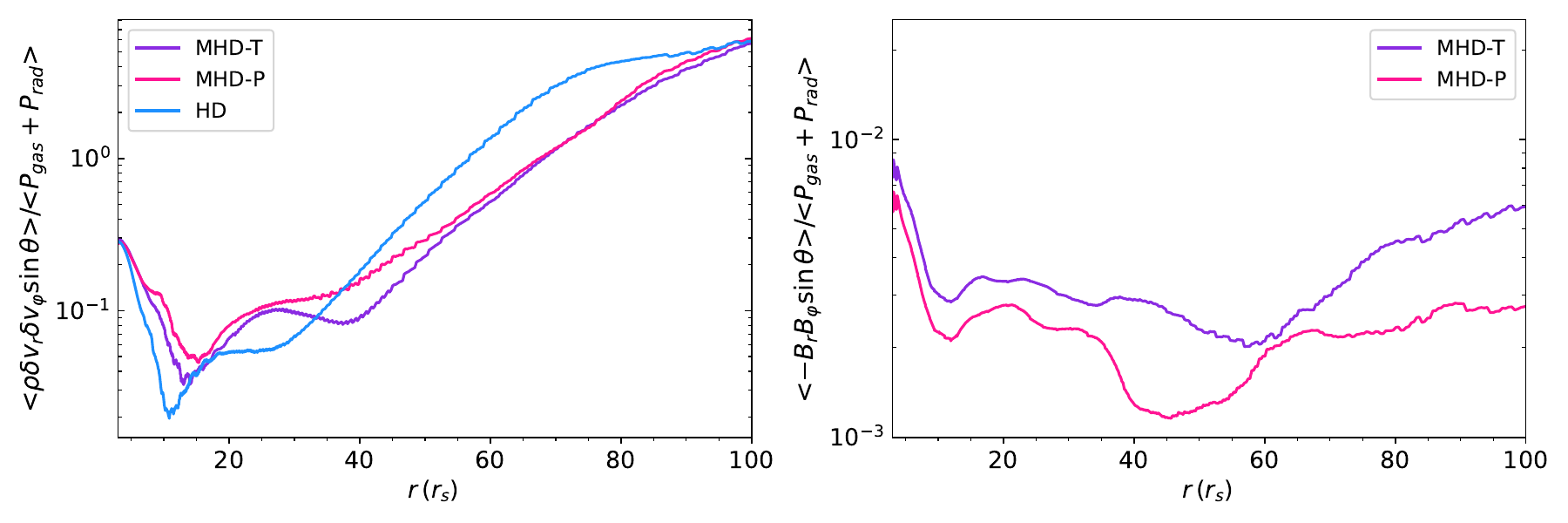}
    \caption{Reynolds (left) and Maxwell (right) stresses averaged within $10$ degrees from the midplane, over azimuth and over the last two days of the simulation, normalized by average of the sum of gas and radiation pressures.}
    \label{fig:5}
\end{figure*}

Figure \ref{fig:5} shows the azimuthally and time-averaged, normalized Reynolds and Maxwell stresses. The rapid growth of the Reynolds stress beyond $\sim25-35\, r_s$ is due to the eccentric infalling stream. Hence, we focus on the values for smaller radii. The Maxwell stress is not significantly affected by the presence of the stream. For radii $r\lesssim25r_s$, the Reynolds stress is larger than the Maxwell stress by factors of $\sim10-50$, for the MHD runs. This suggests that the magnetorotational turbulence is not the main driver of angular momentum transport and, therefore, accretion. Instead, the stream-stream and stream-disk shocks are the main mechanism through which the gas redistributes angular momentum and accretes onto the black hole. This may explain why the accretion rates are broadly similar between the MHD and HD runs after $t\sim1.7$ days. The Reynolds stresses across the MHD runs follow similar radial profiles, while HD shows a radial profile with slightly smaller values for $r\lesssim25r_s$. Comparing the Maxwell stresses between MHD-T and MHD-P shows that the former is between $\sim1.1-2.3$ times larger. The normalized Reynolds stresses vary from $\alpha_{Re}\sim0.02-0.3$ for $r\lesssim 25r_s$. The normalized Maxwell stresses are mostly between $\alpha_{M}\sim0.002-0.006$, consistent with the lower values found for eccentric disks \cite{Chan2024}. Importantly, the normalized Reynolds stress computed here represents angular momentum redistribution resulting from both stream-disk interactions and hydrodynamic turbulence. Although radiation stress can also transport angular momentum, we verified that the radiation viscosity \citep{Loeb1992} is small, as the mean free path of photons is very small in regions where the radiation energy density is large.  We discuss angular momentum redistribution in more detail in Section \ref{ss:angmom}.

\begin{figure*}
    \centering
    \includegraphics[width=\textwidth]{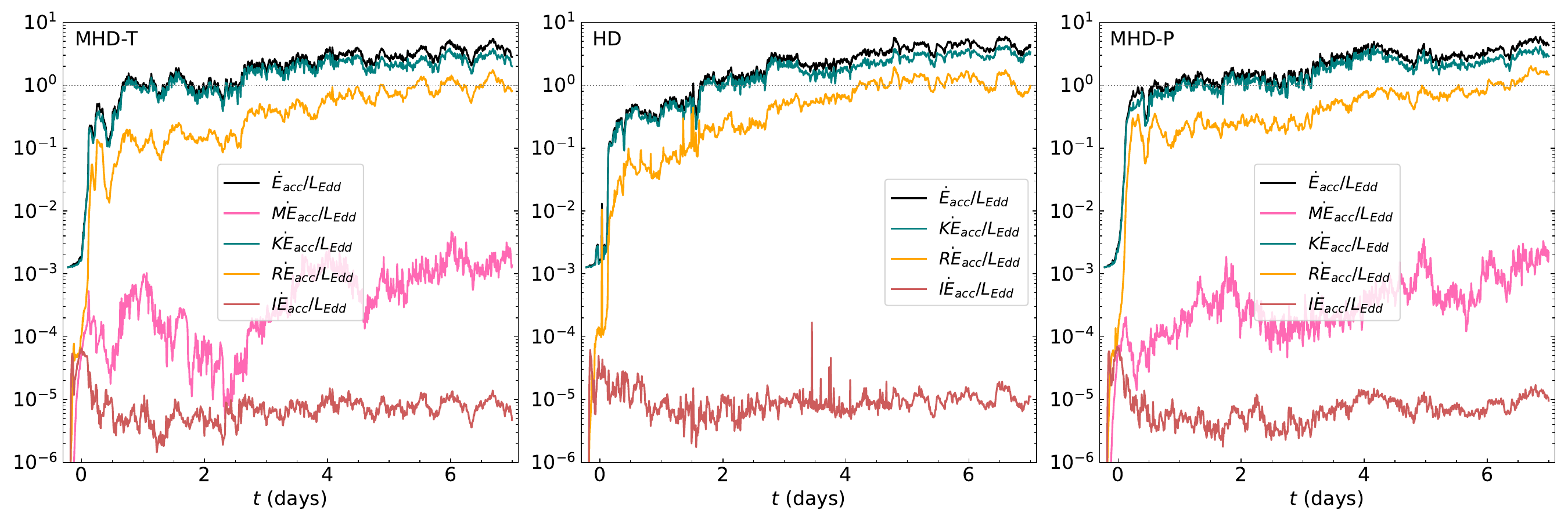}
    \caption{Total accreted energy per unit time (black) and its components: kinetic energy (green), radiative energy (orange), magnetic energy (pink), and internal energy (red) for MHD-T, HD, and MHD-P, from left to right.}
    \label{fig:6}
\end{figure*}

We also compute the total accreted energy per unit time and its components. Figure \ref{fig:6} shows the total, kinetic, internal, magnetic, and radiative energies, in units of the Eddington luminosity $L_{Edd}=4.41\times10^{44}$ erg s$^{-1}$, accreted onto the black hole ($r=3r_s$). The energies per unit time $\dot{E}$ are computed by integrating the energy fluxes over the spherical shell of radius $r$
\begin{equation}
    \dot{E}=\int_A ev_r r^2\sin\theta d\theta d\varphi,
    \label{eq:3}
\end{equation}
where $e$ is the corresponding energy density (i.e., kinetic, internal, or magnetic). The radiative energy per unit time is computed from the lab-frame radial radiation flux, which at $r=3r_s$ is dominated by advection.
The majority of energy accreted is kinetic. The kinetic energy curves mirror the mass accretion curves in Figure \ref{fig:4}. The radiative energy accreted is mostly sub-Eddington, except near the end of the simulations, $t\sim5-6$ days, when the three systems reach slightly super-Eddington values $\dot{RE}_{acc}\sim1-1.5\,L_{Edd}$. By the end of the simulations, this component represents $\sim25\% -35\%$ of the total accreted energy. The internal and magnetic energies accreted are very small in comparison.

\subsection{Outflows}
\label{ss:outflows}

\begin{figure}
    \centering
    \includegraphics[width=0.5\textwidth]{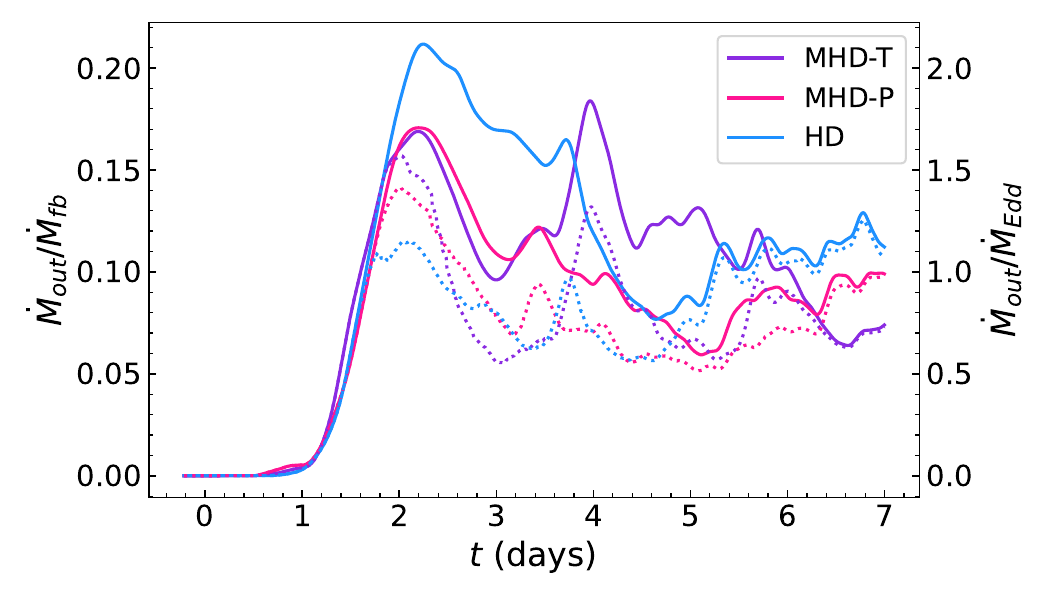}
    \caption{Total mass outflow rate measured at $r=400r_s$ as a function of time since the first stream self-intersection. The dotted lines show the unbound gas.}
    \label{fig:7}
\end{figure}

During the self-crossing and circularization shocks, some gas is accelerated into radiation-driven outflows. We compute the mass outflow rate at our domain boundary $r=400r_s$ using Equation (\ref{eq:2}). The mass outflow rate as a function of time is plotted in Figure \ref{fig:7}. After the first collision shocks, HD has a higher mass outflow rate than the MHD runs. This is due to the different stream thickness at the collision zone, as previously described in \S\ref{s:results}. The maximum outflow rates are initially $\sim17\%$ and $\sim21\%$ of the fallback rate for the MHD runs and HD, respectively. The periodicity observed roughly reflects the stream self-interactions mixed with the interactions of different outflowing shells of gas as they expand. As the stream-stream and stream-disk collisions get weaker, less gas is accelerated into outflows. The decrease in the mass outflow rate is steeper for HD, while in the MHD runs, it is more gradual. By the end of the simulations, the outflow rate is $\sim7-12\%$ of the fallback rate. 

Figure \ref{fig:7} also shows the mass outflow rate through $r=400r_s$ of the unbound gas as dotted lines. Similar to \cite{JiangStoneDavis2014}, we define the unbound gas as that which has a positive Bernoulli parameter given by
\begin{equation}
    Be=\rho\Phi_{TR}+\frac{1}{2}\rho v^2+\frac{\gamma P_{gas}}{\gamma-1}+\frac{4}{3}E_{rad},
    \label{eq:bernoulli}
\end{equation}
where the terms are the gravitational potential energy density, the kinetic energy density, the gas enthalpy density ($\gamma=5/3$), and the sum of the radiation energy density and the radiation pressure.

After the first stream-stream collisions, between $t\sim0.5-1.7$ days, the unbound gas represents $100\%$ of the total outflow rate. These unbound outflows have mass-averaged speeds in the range $\sim0.08-0.18c$ measured at $r=400r_s$. As the collisions continue, the acceleration of outflows decreases and the unbound fraction decreases to $\sim40-80\%$. After $t\sim3.3$ days, the unbound outflow has speeds $\sim0.05-0.10c$ measured at $r=400\,r_s$. By the end of the simulations, around $\sim94-99\%$ of the gas that reaches $r=400r_s$ is unbound. At smaller radii $r\lesssim300 r_s$, a significant fraction of gas that was initially launched is falling back toward the black hole a few days after the first stream-stream collision. The mass-averaged outflow speeds of the bound gas generally decrease with time and vary between $\sim0.005-0.05c$ at $r=400r_s$.

\begin{figure*}
    \centering
    \includegraphics[width=\textwidth]{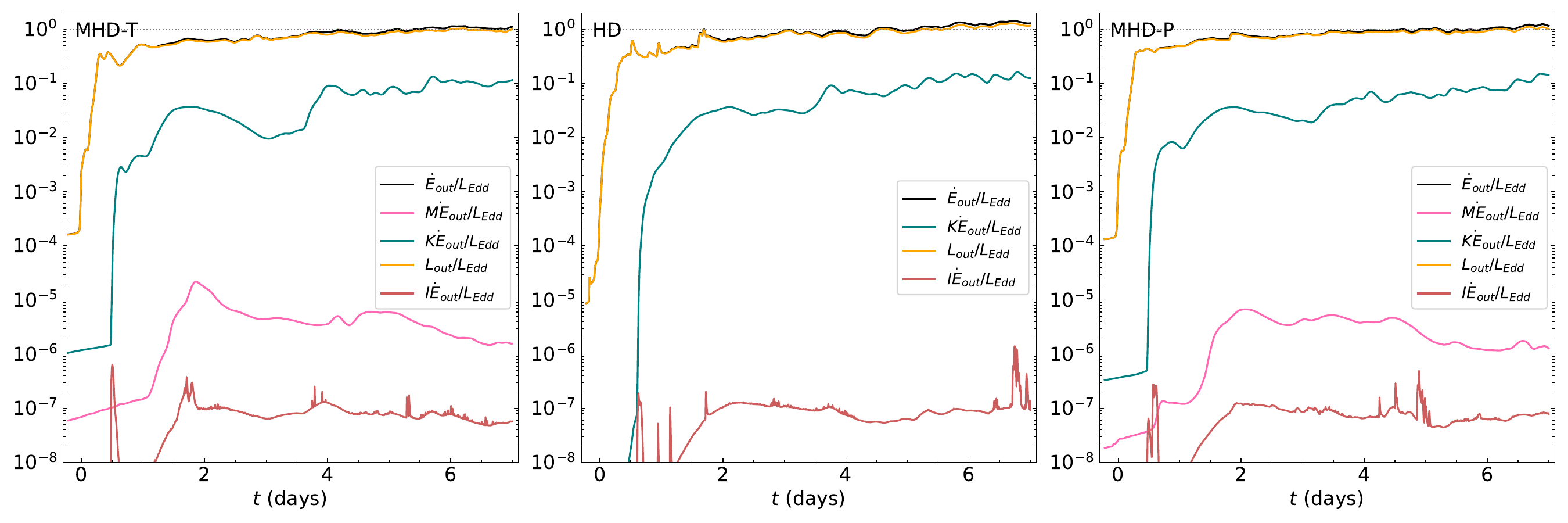}
    \caption{Total outflowing energy per unit time (black) at $r=400r_s$ and its components: kinetic energy (green), luminosity (orange), magnetic energy (pink), and internal energy (red) for MHD-T, HD, and MHD-P, from left to right.}
    \label{fig:8}
\end{figure*}

The energy content of the outflows is computed as in Equation (\ref{eq:3}) at $r=400r_s$ and is shown in Figure \ref{fig:8}. The radiative luminosity accounts for most of the outflowing energy, as expected in radiation-driven outflows. The outflowing kinetic energy is between $\sim3-15\%$ that of the radiation. The magnetic and internal energies in the outflows are quite small, consistent with optically thick outflows that are neither magnetically nor thermally accelerated.

Comparing Figures \ref{fig:4} and \ref{fig:7}, shows that after the first stream-stream collisions, outflows dominate by factors between $\sim1.2-2$ over the mass accretion rate. Around $t\sim3$ days, the accretion rate increases above the outflow rate. By the end of the simulations, the accretion rate is 3-6 times larger than the outflow rate. 

\begin{figure*}
    \centering
    \includegraphics[width=\textwidth]{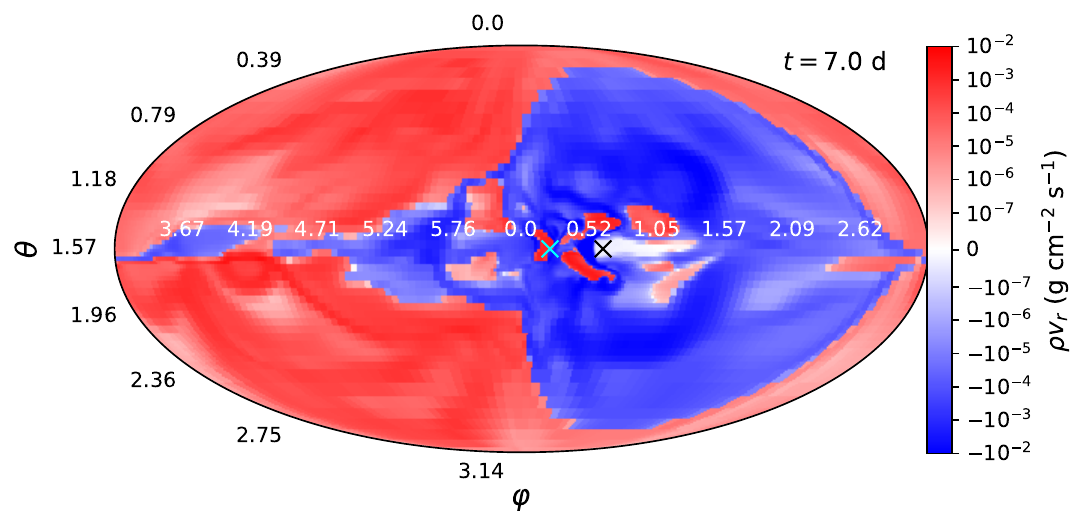}
    \caption{Mollweide projection of the radial mass flux for MHD-T measured at $r=100r_s$ and $t=7$ days. The cyan $\times$ marks the injection angle at $r=400r_s$, and the black $\times$ marks the first stream-stream collision angle. The projection is centered at $\varphi=0$ and is shown in terms of the spherical polar coordinates $\theta$ and $\varphi$.}
    \label{fig:mollweide}
\end{figure*}

During the first stream-stream collision, gas is initially launched from the midplane at the collision zone $r\sim60-80 r_s$ from the black hole. The collision zone thereafter is spread over a wider range of radii and azimuthal angles as the stream interacts with the shocked gas that accumulates near the black hole. Figure \ref{fig:mollweide} shows the Mollweide projection of the radial mass flux measured at $r=100r_s$ at the end of the MHD-T run. We can see that gas is mostly outflowing at angles $5\pi/6\lesssim\varphi\lesssim2\pi$ for most polar angles. Although this plot includes both bound and unbound gas, we verified that unbound outflows follow a similar pattern and are therefore preferentially found in this azimuthal range. In the range $0\lesssim\varphi\lesssim5\pi/6$, gas is inflowing at most polar angles. In addition, very close to the midplane, the inflow of gas occurs at most azimuthal angles. This general azimuthal and polar angle flow structure is established at $t\sim3$ days at $r=100r_s$, and can be seen at larger radii $r\sim350r_s$ at late times as the gas launched interacts with the fallback stream. The outflow is therefore largely non-axisymmetric, with a structure set by the stream-disk collisions and the interactions of launched gas with the stream at larger radii.

\subsection{Radiation}
\label{ss:radiation}

\begin{figure}[h!]
    \centering
    \includegraphics[width=0.5\textwidth]{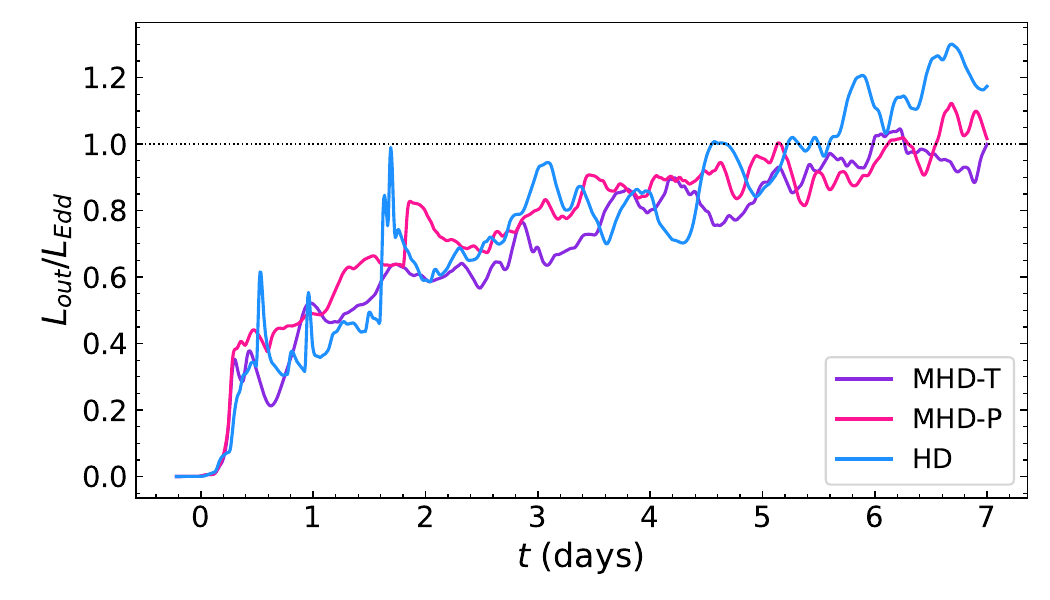}
    \caption{Luminosity measured at $r=400r_s$ in units of the Eddington luminosity as a function of time since the first self-intersection collision.}
    \label{fig:10}
\end{figure}

Figure \ref{fig:10} compares the evolution of the radial luminosity of the three simulations. The luminosity tends to increase with time as more gas shocks, dissipating orbital energy into radiation. HD shows relatively sharper peaks at early times compared to the MHD runs, while the stream-stream collisions are still happening. This may indicate different diffusion timescales through the collision zone and outflows. Otherwise, their luminosity profiles are not significantly distinct. The luminosity increases rapidly after the first stream-stream collision, and then gradually increases to $\sim4-6\times10^{44}$ erg s$^{-1}$ by the end of the simulations. This corresponds to slightly super-Eddington luminosities $L/L_{Edd}\sim1-1.3$. We find that the effective photosphere, considering the Planck mean absorption and electron scattering opacities, is highly dependent on viewing angle. We measure the luminosity at our domain boundary, but given that the outflows remain optically thick throughout, the radiation may still be reprocessed at larger radii. The rise in luminosity in our simulations may be artificially quick given our constant super-Eddington mass fallback rate.

\begin{figure*}
    \centering
    \includegraphics[width=\linewidth]{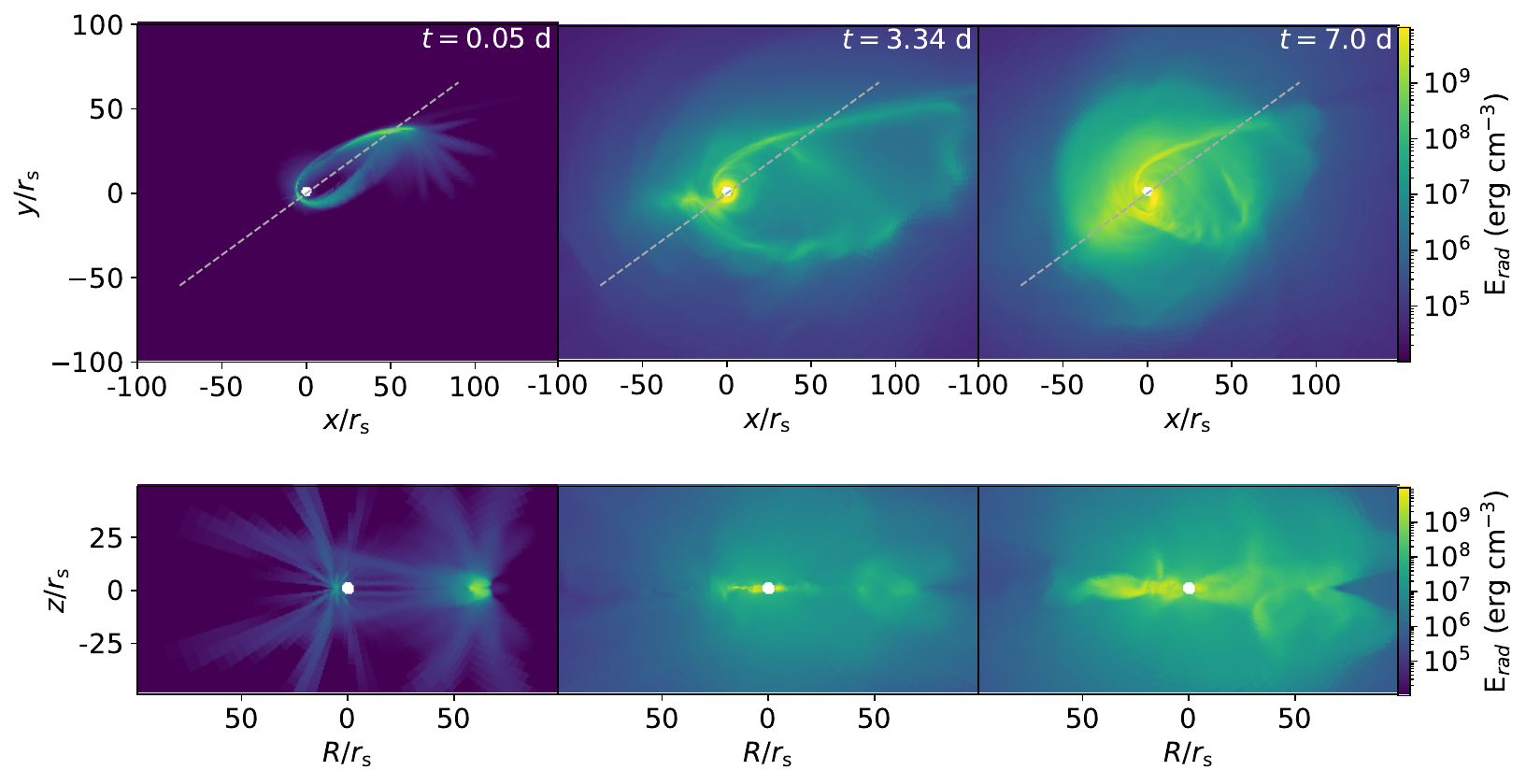}
    \caption{Lab-frame radiation energy density slices of MHD-T during the initial self-intersection shock, around the middle of the simulation, and the final snapshot.
    The top row shows midplane slices, while the bottom row shows vertical slices at the collision angle (shown as a dashed line in the top row panels).}
    \label{fig:er}
\end{figure*}

Figure \ref{fig:er} shows radiation energy density slices of MHD-T, during the first stream-stream collision, around the middle of the simulation, and at the end of the run. The first panel shows that radiation is initially emitted predominantly in the stream-stream collision zone, although some emission at the pericenter is observed as well, due to the strong compression. As the simulation progresses, the interacting length of the stream remains bright, and some dissipation also occurs at inner radii as the stream encounters the accumulating material near pericenter. The last panel shows emission from shocks and a growing accretion component. 

We estimate the radiative efficiency in two ways
\begin{equation}
    \begin{aligned}
        &\eta_{acc}=\frac{L_{out}}{\dot{M}_{acc}c^2},\\
        &\eta_{fb}=\frac{L_{out}}{\dot{M}_{fb}c^2},
    \end{aligned}
    \label{eq:efficiencies}
\end{equation}
where $\eta_{acc}$ and $\eta_{fb}$ are the radiative efficiencies with respect to the mass accretion rate and the mass fallback rate, respectively. We find that initially, on average, $\eta_{acc}\sim5\%$ for the MHD runs, with a higher average efficiency $\eta_{acc}\sim6.5\%$ for HD. By $t\sim3-4$ days, the radiative efficiencies have decreased to $\eta_{acc}\sim3\%$ for the three runs. This occurs concurrently with the decrease in the outflow rate and increase in the accretion rate (see Figures \ref{fig:4}, \ref{fig:7}). For accretion emission we can expect $L\propto \dot{M}_{acc}$; however, the increase in the accretion rate did not result in a scaled increase in the luminosity in these systems. This implies that although there may be a luminosity component resulting from accretion, it is not dominant at this stage (see Figure \ref{fig:er}). In comparison, after the first stream-stream collision $\eta_{fb}\sim0.4\%$, thereafter following the growth of the luminosity (see Figure \ref{fig:10}). By the end of the simulations $\eta_{fb}\sim1-1.3\%$. We find that $\kappa_s>\kappa_a$ throughout most of the domain; the scattering opacity then sets the Eddington luminosity. The radiative efficiency is then Eddington limited, as excess radiative energy is expended in the acceleration of outflows.

Additionally, we compute the kinetic efficiency defined as 
\begin{equation}
    \eta_{ke}=\frac{\dot{KE}_{out}}{{\dot{M}_{fb}c^2}},
    \label{eq:keeff}
\end{equation}
where $\dot{KE}_{out}$ is kinetic energy per unit time outflowing at $r=400r_s$. This efficiency reaches $\sim0.04\%$ after the first stream self-intersection and increases gradually to $\sim0.1-0.16\%$ by the end of the runs. The time-averaged efficiencies are summarized in Table \ref{tab:1}.

\begin{table}
\centering
 \begin{tabular}{c| c| c| c| c} 
 
 Simulation & $\dot{M}_{acc}/\dot{M}_{Edd}$ & $\eta_{acc}$ & $\eta_{fb}$& $\eta_{ke}$  \\ [0.5ex] 
 \hline 
 MHD-T & $3.54$ & $0.028$& $0.0094$ & $0.00096$\\
 MHD-P & $3.60$ &$0.028$& $0.0095$ & $0.00088$\\
 HD & $4.22$ &$0.027$& $0.011$ & $0.0012$ \\ [1ex] 
 \end{tabular}
 \caption{\label{tab:1} Mass acretion rate, radiative efficiency with respect to the accretion and fallback rates (Equation \ref{eq:efficiencies}), and kinetic energy efficiency (Equation \ref{eq:keeff}) averaged over the last two days of the simulations. Note that in the ratio $\dot{M}_{acc}/\dot{M}_{Edd}$, $\dot{M}_{Edd}$ is computed assuming a radiative efficiency $\eta=0.1$.}
\end{table}

\subsection{Angular momentum redistribution}
\label{ss:angmom}
In this section, we use the equation of conservation of angular momentum to more completely characterize the angular momentum redistribution throughout the forming accretion flow.

The momentum equation in its conservative form is
\begin{equation}
    \frac{\partial}{\partial t}(\rho\bm{v})+\nabla\cdot\big(\rho\bm{vv}+P\tilde{I}-\bm{BB}\big)=\bm{f}_{ext},
    \label{eq:a1}
\end{equation}
where $\bm{v}$ is the flow velocity, $P=P_{gas}+P_{mag}$ is the total pressure, $\bm{B}$ is the magnetic field, $\tilde{I}$ is the identity matrix and $\bm{f}_{ext}$ is the sum of external forces per unit volume.
The azimuthal component of Equation (\ref{eq:a1}) expresses angular momentum conservation. In spherical polar coordinates this can be written as
\begin{equation}
\begin{aligned}
    &\frac{\partial}{\partial t}(\rho v_{\varphi})+\frac{1}{r^3}\frac{\partial}{\partial r}\Big[r^3\big(\rho v_rv_{\varphi}-B_rB_{\varphi}\big)\Big]\\
    &+\frac{1}{r\sin^2\theta}\frac{\partial}{\partial\theta}\Big[\sin^2\theta\big(\rho v_{\theta}v_{\varphi}-B_{\theta}B_{\varphi}\big)\Big]\\
    &+\frac{1}{r\sin\theta}\frac{\partial}{\partial\varphi}\big(\rho v_{\varphi}^2+P-B_{\varphi}^2\big)=f_{ext,\varphi}.
    \label{eq:a2}
\end{aligned}
\end{equation}

We then average over azimuthal angle. Since we are interested in the accretion flow forming near the midplane, we multiply by $\sin\theta$ and average over $\theta$ from $\theta_{-}=\pi/2-\theta_0$ to $\theta_{+}=\pi/2+\theta_0$, where $\theta_0$ is a small angle from the midplane, which we choose to be $\theta_0=10$ degrees. This yields

\begin{equation}
\begin{aligned}
    &\frac{\partial}{\partial t}\big(\langle\langle r^3\rho v_{\varphi}\sin\theta\rangle\rangle\big)+\frac{\partial}{\partial r}\big(r^3\langle\langle \sin\theta(\rho v_rv_{\varphi}-B_rB_{\varphi})\rangle\rangle\big)\\
    &+\frac{r^2\cos^2\theta_0}{2\sin\theta_0}\big(\langle\rho v_{\theta}v_{\varphi}-B_{\theta}B_{\varphi}\rangle|_{\theta_{+}}-\langle\rho v_{\theta}v_{\varphi}-B_{\theta}B_{\varphi}\rangle|_{\theta_{-}}\big)\\
    &=r^3\langle\langle\sin\theta f_{ext,\varphi}\rangle\rangle,
    \label{eq:a4}
\end{aligned}
\end{equation}
where we have also multiplied through by $r^3$. Here, $\langle X\rangle$ represents the azimuthal average of variable $X$,  and $\langle\langle X \rangle\rangle$ denotes the azimuthal and polar average.

Now, consider the stress term $\rho v_rv_{\varphi}\sin\theta$ inside the $\partial/\partial r$ term. We can decompose it into mean and fluctuating components
\begin{equation}
    \rho v_r=\langle\langle\rho v_r\rangle\rangle+\delta(\rho v_r),
    \label{eq:a5}
\end{equation}

\begin{equation}
    v_{\varphi}\sin\theta=\langle\langle v_{\varphi}\sin\theta\rangle\rangle+\delta(v_{\varphi}\sin\theta).
    \label{eq:a6}
\end{equation}

Therefore,
\begin{equation}
    \langle\langle\rho v_rv_{\varphi}\sin\theta\rangle\rangle=\langle\langle\rho v_r\rangle\rangle\langle\langle v_{\varphi}\sin\theta\rangle\rangle+\langle\langle\delta(\rho v_r)\delta(v_{\varphi}\sin\theta)\rangle\rangle.
    \label{eq:a7}
\end{equation}
We can connect the mean term in Equation (\ref{eq:a7}) with the mass accretion rate and the specific angular momentum through the disk of height $2r\sin\theta_0$ and circumference $2\pi r$
\begin{equation}
    \dot{M}=-4\pi r^2\sin\theta_0\langle\langle\rho v_r\rangle\rangle,
    \label{eq:a8}
\end{equation}
\begin{equation}
    l=r\langle\langle v_{\varphi}\sin\theta\rangle\rangle.
\end{equation}
Equation (\ref{eq:a4}) can then be written as

\begin{equation}
    \begin{aligned}
        &\frac{\partial}{\partial t}\big(\langle\langle r^3\rho v_{\varphi}\sin\theta\rangle\rangle\big)+\frac{\partial}{\partial r}\bigg[\frac{-\dot{Ml}}{4\pi\sin\theta_0}+ r^3\langle\langle \delta(\rho v_r)\delta(v_{\varphi}\sin\theta)\\
        &-B_rB_{\varphi}\sin\theta\rangle\rangle\bigg]
        +\frac{r^2\cos^2\theta_0}{2\sin\theta_0}\big(\langle\rho v_{\theta}v_{\varphi}-B_{\theta}B_{\varphi}\rangle|^{\theta_{+}}_{\theta_{-}}\big)\\
        &=r^3\langle\langle\sin\theta f_{ext,\varphi}\rangle\rangle.
    \label{eq:a10}
    \end{aligned}
\end{equation}

In standard accretion disk theory, it is often assumed that the disk is in or near a steady state, and that the vertical fluxes through the surface of the disk and the torques exerted by external forces are negligible. These assumptions would imply
\begin{equation}
    \dot{M}l=4\pi\sin\theta_0r^3\langle\langle \delta(\rho v_r)\delta(v_{\varphi}\sin\theta)-B_rB_{\varphi}\sin\theta\rangle\rangle + C,
    \label{eq:a11}
\end{equation}
where $C$ is a constant of integration set by the radial boundary conditions.
Equation (\ref{eq:a11}) states that mass accretion is driven by the transport of specific angular momentum by the turbulent Reynolds and Maxwell stresses. This picture may be significantly altered when the other terms in Equation (\ref{eq:a10}) are not negligible.

\begin{figure}[h!]
    \centering
    \includegraphics[width=0.5\textwidth]{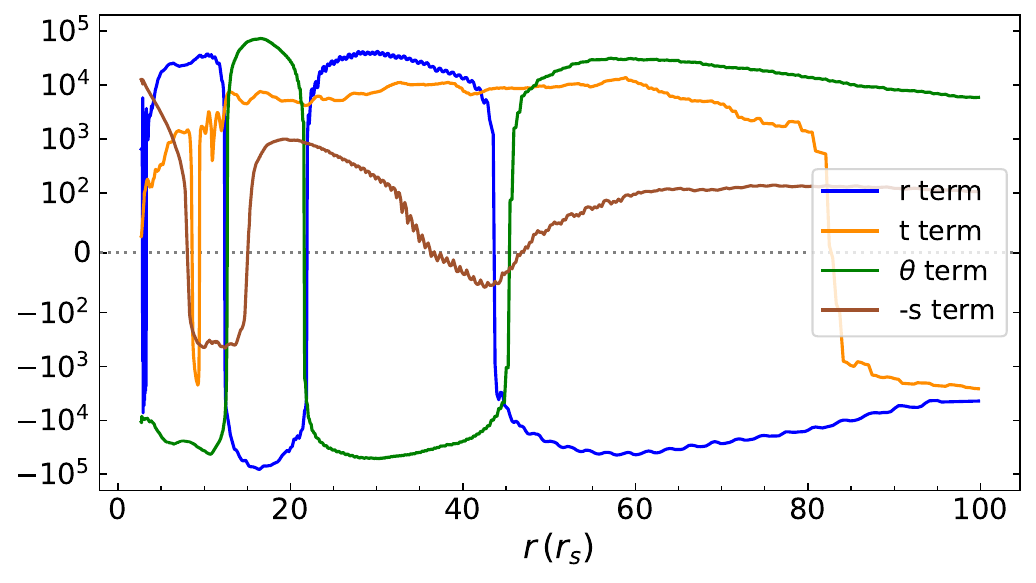}
    \caption{Terms in the angular momentum conservation equation (\ref{eq:a10}) for MHD-T in code units. The terms are averaged over the last two days of the simulation in azimuth and 10 degrees from the midplane.}
    \label{fig:a1}
\end{figure}

Figure \ref{fig:a1} shows the individual terms in Equation (\ref{eq:a10}) for MHD-T averaged over the last two days of the simulation. We label the terms as the time, radial, $\theta$, and source terms, respectively. Notably, the radial and $\theta$ terms show a strong anti-correlation. The $\theta$-term in Equation (\ref{eq:a10}), being dominated by the $\rho v_{\theta}v_{\varphi}$ terms, is positive when vertical outflows are large and negative when gas is mostly inflowing through the surfaces of the disk. We find that vertical torques from the magnetic field are small, as mediated by the $B_{\theta}B_{\varphi}$ component in the $\theta$-term. This suggests that the magnetic launching of outflows is not significant. The time term is large, indicating that the system is far from a steady state. In this system, the source term consists of the sum of the radiation and gravitational forces. We found that the former is very small and that the latter dominates this term. This gravitational acceleration in the azimuthal direction is a crucial feature of the generalized Newtonian potential \citep{Tejeda2013} to reproduce relativistic apsidal precession. Meanwhile, radiation torques play a negligible role in angular momentum transport.

To interpret how these terms modulate the accretion rate directly, we integrate Equation (\ref{eq:a10}) from an inner radius $r_{in}=3r_s$ to $r$ and solve for $\dot{M}$ to get

\begin{equation}
    \begin{aligned}
        \dot{M}&=\frac{4\pi\sin\theta_0}{l}\bigg[r^3\langle\langle \delta(\rho v_r)\delta(v_{\varphi}\sin\theta)-B_rB_{\varphi}\sin\theta\rangle\rangle\\
        &+\int^r_{r_{in}}\frac{r^2\cos^2\theta_0}{2\sin\theta_0}\langle\rho v_{\theta}v_{\varphi}-B_{\theta}B_{\varphi}\rangle|^{\theta_{+}}_{\theta_{\_}}dr\\
        &+\int^r_{r_{in}}\frac{\partial}{\partial t}\langle\langle r^3\rho v_{\varphi}\sin\theta\rangle\rangle dr\\
        &-\int^r_{r_{in}}r^3\langle\langle\sin\theta f_{ext,\varphi}\rangle\rangle dr- C\bigg],
        \label{eq:a12}
    \end{aligned}
\end{equation}

where
\begin{equation}
C=\Big[-\frac{\dot{Ml}}{4\pi\sin\theta_0}+r^3\langle\langle \delta(\rho v_r)\delta(v_{\varphi}\sin\theta)- B_rB_{\varphi}\sin\theta\rangle\rangle\Big]_{r_{in}}
\end{equation}
is the integrated radial term evaluated at the inner radius.

\begin{figure}[h!]
    \centering
    \includegraphics[width=0.5\textwidth]{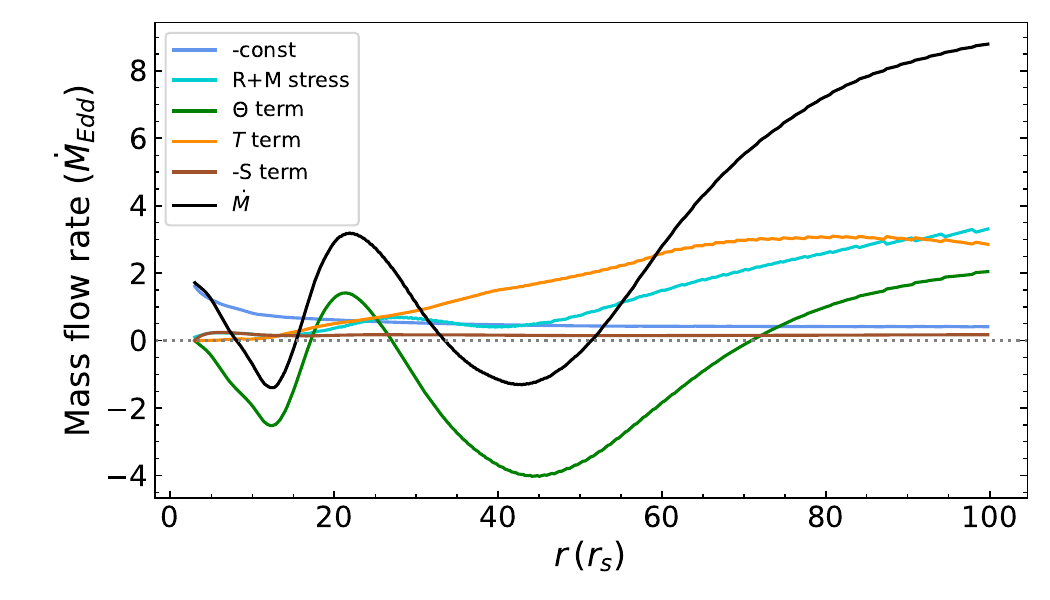}
    \caption{Terms in the integrated angular momentum conservation equation (\ref{eq:a12}) for MHD-T in units of $\dot{M}_{Edd}$. The terms are averaged over the last two days of the simulation in azimuth and 10 degrees from the midplane.}
    \label{fig:a2}
\end{figure}

These terms are plotted in Figure \ref{fig:a2} and are labeled as the R+M (Reynolds + Maxwell) stress term, the $\Theta$ term, the $T$ term, the $S$ term and the constant term, respectively. This plot, along with Figure \ref{fig:a1}, shows how it is mainly the outflows and inflows through the surface of the disk that modulate the accretion rate. Vertical angular momentum fluxes balance radial angular momentum fluxes, which are then reflected in the mass accretion rate. The R+M stress term is dominated by the Reynolds stress. These internal torques contribute positively to the accretion rate, in a moderate way. The time term shows that mass is accumulating with time in the disk. The source term has the least influence on the accretion rate. This analysis shows that the Reynolds and Maxwell stresses alone do not provide a complete characterization of the mass accretion in this highly dynamical system.

\subsection{Accretion flow structure}
\label{ss:structure}
\subsubsection{Eccentricity}
\label{sss:eccentricity}
\begin{figure}[h!]
    \centering
    \includegraphics[width=0.5\textwidth]{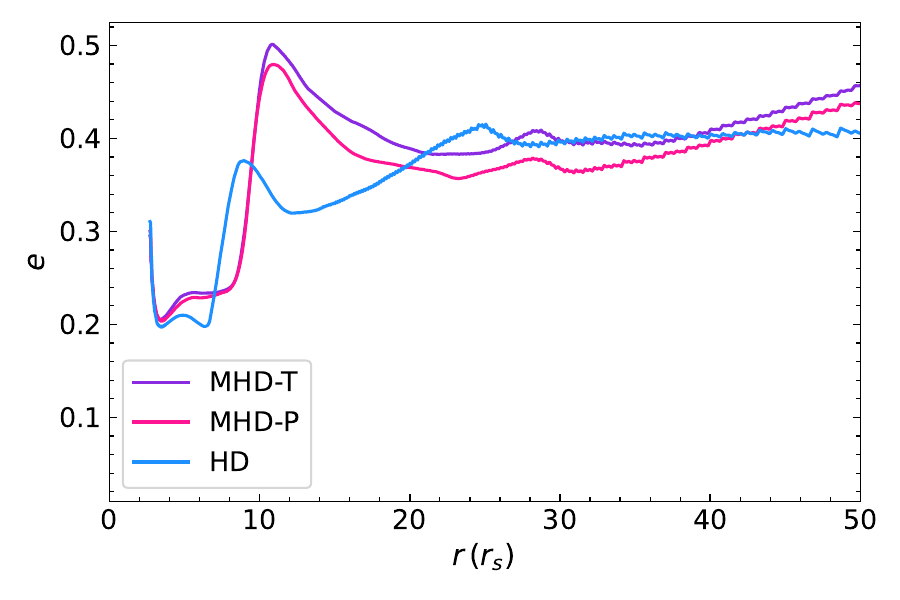}
    \caption{Mass-weighted eccentricity averaged within $10$ degrees from the midplane, over azimuth and over the last two days of the simulation.}
    \label{fig:11}
\end{figure}

To measure the level of circularization of the accretion flow, we estimate the eccentricity as in \cite{Oyang2021}. We compute the mass-weighted, azimuthally-averaged eccentricity for gas within $10$ degrees of the midplane. Since we inject the debris stream at a constant rate throughout the duration of the simulation, we reduced the impact of the stream for this eccentricity measurement through density and radial and azimuthal velocity cuts for $r\gtrsim20 \,r_{s}$. Inside $r\lesssim20 \,r_{s}$, it becomes difficult to differentiate the stream from the disk. We averaged the radial dependence of the eccentricity over the last two days of the simulation; this is shown in Figure \ref{fig:11}. Close to the black hole $r\lesssim\,8r_{s}$, the accretion flow is moderately circularized with $e\sim0.2-0.3$, while for $10\lesssim r \,(r_s)\lesssim50$ the forming disk remains eccentric with $e\sim0.4-0.5$. This results from the stream continually depositing high eccentricity gas into the accretion flow. Although the MHD runs are moderately circularized out to $\sim2 \,r_s$ further from the black hole compared to HD, for the bulk of the accretion flow, they tend to have similar eccentricities.

\subsubsection{Azimuthal and polar angle dependence}
\label{sss:azimuthaldep}
\begin{figure}
    \centering
    \includegraphics[width=0.5\textwidth]{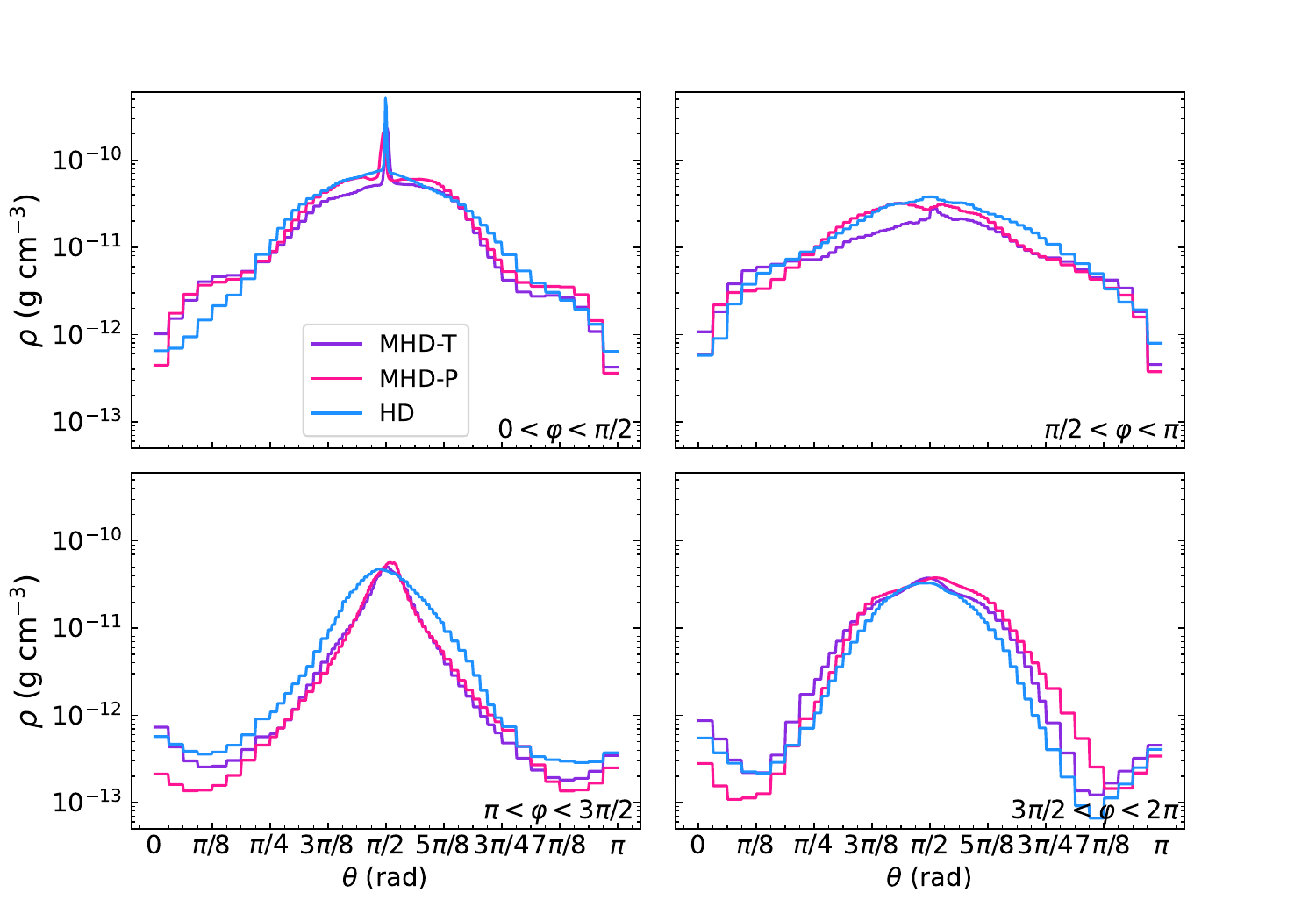}
    \caption{Density at $r=40 \,r_s$ for four azimuthal bins as a function of polar angle. The profiles shown are time-averaged over the last two days of the simulation.}
    \label{fig:12}
\end{figure}

We now examine the structure of the accretion flow as a function of azimuth and polar angle. We first compute the density profiles at $r=40 \,r_s$ as a function of polar angle averaged in four azimuthal bins, as shown in Figure \ref{fig:12}. This radius is chosen because the majority of the accretion flow is contained within $\sim50\,r_s$. We average over the last two days of the simulation; however, we note that the density structure is still noticeably dynamic for regions $\gtrsim\pi/8$ from the midplane.  The first azimuthal bin ($0-\pi/2$), is the quadrant where the stream-stream and stream-disk collisions primarily occur. The sharp peak at the midplane is the dense stream, which is densest for HD given the lack of magnetic pressure support. In all quadrants, the density profile peaks near the midplane and mostly decreases towards the poles. Additionally, there is an azimuthal dependence to the density structure of the accretion flow. In the collision quadrant and the adjacent one ($\pi/2-\pi$) the density profiles decrease the slowest, maintaining higher densities further from the midplane. By producing a similar plot for $v_{\theta}$, we find that in the collision quadrant gas is mostly outflowing from the midplane, and mostly falling back in the adjacent quadrant ($\pi/2-\pi$), which results in broadened density profiles. In contrast, the quadrant opposite ($\pi-3\pi/2$) to the collision quadrant shows the steepest decrease in density with distance from the midplane as the gas is interacting less in that region. This additional azimuthal dependence on the density structure will result in an azimuthal dependence on the optical depth, where near the collision zone, larger optical depths may be expected even further from the midplane compared to those for which the line of sight is opposite to the collision zone. 

We find that the Planck mean photosphere (where $\tau_P=\int \rho\kappa_P dr=1$) for all runs is mostly outside the grid boundaries due to large atomic opacities. The outflows are optically thick and the temperatures measured at $r=400\,r_s$ are on average $T\sim3-5\times10^4$ K, which is consistent with an optical/UV source. Sustained X-ray emission for some viewing angles may occur after the outflows subside; this is being investigated in a companion paper (Huang et al. in prep). 

\begin{figure}
    \centering
    \includegraphics[width=0.5\textwidth]{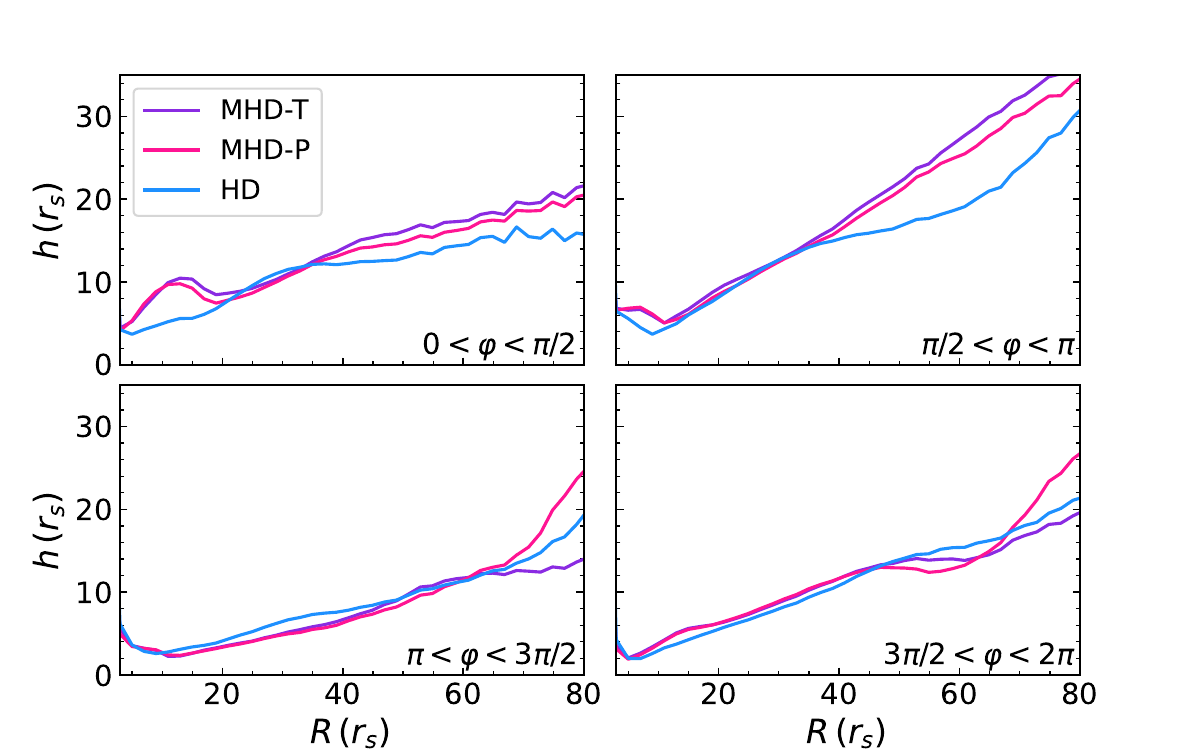}
    \caption{Disk thickness computed for four azimuthal bins as a function of cylindrical radius. The profiles shown are time-averaged over the last two days of the simulation.}
    \label{fig:13}
\end{figure}

\begin{figure}
    \centering
    \includegraphics[width=0.4\textwidth]{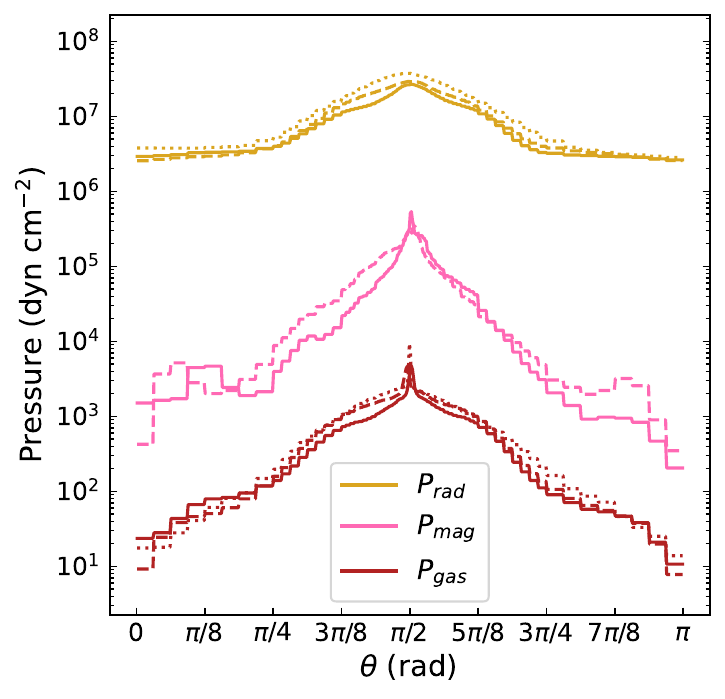}
    \caption{Azimuthally-averaged radiation ($P_{rad}$), magnetic ($P_{mag}$), and gas ($P_{gas}$) pressures as a function of polar angle at $r=40 r_s$, averaged over the last two days of the simulations. The profiles for MHD-T, MHD-P, and HD are shown as solid, dashed, and dotted lines, respectively.}
    \label{fig:press}
\end{figure}

Next, we compute the time-averaged disk thickness as a function of cylindrical radius $R$ averaged over four azimuthal bins, shown in Figure \ref{fig:13}. We estimate the disk thickness $h$ through the first moment of the density
\begin{equation}
    h=\frac{\sum\limits_i \rho_i |z_i|dV_i}{\sum\limits_i\rho_i dV_i},
    \label{eq:scaleheight}
\end{equation}
where $\rho$ is the density, $z$ is the cylindrical coordinate, $dV$ is the volume of the cell and the summation is done over the $i$th cylindrical radius bin.
 In the collision quadrant, $h$ is slightly larger for the MHD runs. This larger vertical extent can also be seen on the right side of the vertical slice panels in Figure \ref{fig:3}, which cut through the collision angle. Note that this is not due to additional magnetic pressure support in the disk (see Figure \ref{fig:press}), but rather to the different thicknesses of the infalling stream. Given that the stream is more diffuse in the MHD runs, as it intersects the accretion flow, more of the stream gas is redistributed from the midplane to larger angles, resulting in a slightly larger estimate of the disk thickness. This lofted gas is not in hydrostatic equilibrium. The quadrant opposite to the collision quadrant generally has the shortest vertical extent, by a factor that mostly varies between $\sim0.5-0.7$ that of the collision quadrant. This is consistent with the density structure found for the ($\pi-3\pi/2$) bin, where the density profile is narrower around the midplane. The aspect ratio varies with both azimuth and radius and is largely between $h/r\sim0.2-0.5$.

 Figure \ref{fig:press} shows the azimuthally-averaged radiation, magnetic, and gas pressures as a function of polar angle measured at $r=40r_s$, averaged over the last two days of the simulation. Throughout polar angle, the radiation pressure is much higher than the magnetic and gas pressures. Near the midplane, the radiation pressure is $\sim1-2$ and $\sim3-4$ orders of magnitude larger than the magnetic and gas pressures, respectively. Toward the poles, these factors increase to $\sim3-4$ and $\sim5$ orders of magnitude, respectively. Their gradients show that the radiation force sets the vertical structure of the accretion flow. The radiation pressure continues to dominate by orders of magnitude out to large radii. 

\section{Discussion}
\label{s:discussion}
\subsection{Comparison with other (R)MHD works}
\label{ss:othermhd}

Our results indicate that in the debris fallback stage following a TDE, the main effect of large magnetic fields is to increase the thickness of the stream. In comparison with a purely radiation-hydrodynamic stream, when the stream self-crossing shock happens between the returning and infalling parts of the highly magnetized stream the collision zone is less dense, and therefore less opaque to the radiation produced during the strong shock. This results in a less efficient radiative acceleration of the gas in the collision zone, yielding weaker initial outflows ($\sim0.67\times$ smaller outflow rate) and more initial accretion ($\sim2-6\times$ more) into the black hole. This picture is most pronounced initially, when the collision shocks occur between the infalling and returning streams. As gas accumulates near the black hole and the stream interacts with the forming accretion flow instead, the broad dynamics of the magnetized and nonmagnetized systems becomes largely indistinct. This result is in agreement with previous work that studied weak fields in a different subset of the TDE parameter space.

\cite{Sadowski2016} performed the first general relativistic magnetohydrodynamic simulations of TDEs that studied both the disruption and the accretion flow formation phases. They studied the deep $\beta=r_t/r_p=10$ disruption of a $M_*=0.1M_{\odot}$ red dwarf by a $M_{BH}=10^5M_{\odot}$ using an SPH code with a relativistic treatment to both gravity and hydrodynamics. The debris was then mapped to a \texttt{KORAL} grid, where a weak poloidal magnetic field was first added to the debris. They found that the self-crossing shocks launch thermally-accelerated outflows and quickly form a thick $h/r=2$ eccentric disk. This accretion flow structure is affected by the absence of cooling through radiation in their simulations. In contrast, we find $h/r\sim0.2-0.5$ varying with radius and azimuth. We find similar values for the Reynolds stress close to the black hole $\alpha\sim 0.2$ as their measured effective $\alpha\sim0.4$ at $r\sim 5\,r_s$. In agreement, we also find that the Reynolds stress dominates over the Maxwell stress, although by more than an order of magnitude, as they found. By analyzing the conservation of angular momentum in our simulations, we found that vertical outflows and inflows regulate the accretion rate rather than the turbulent internal torques. Similarly in their simulations at late times, given that their self-interaction stage ends promptly, it is inflow of the initially ejected gas that dominates mass accretion and not turbulent viscosity.

Our approach is more similar to that in \cite{Curd2021} who studied the accretion flow formation stage by injecting the debris stream into a GRRMHD \texttt{KORAL} grid. One of their simulations modeled a stream with a weak poloidal field $\beta_M=100$ resulting from the close disruption $\beta=5$ of a $M_*=1M_{\odot}$ star by a $M_{BH}=10^6M_{\odot}$ black hole in a bound orbit $e=0.97$. In comparison, our parameters more closely resemble those expected in TDEs $\beta=1.73$ and $e=0.99$, but with a much larger magnetic field $\beta_M=0.05$, and an additional toroidal field geometry, to capture the effects of fields. In their simulation, the magnetorotational instability is not resolved; however, they find that the magnetic field increases through the winding of the field, but becomes increasingly subdominant to the total gas and radiation pressure. Due to our larger initial magnetic fields and finer resolution, we may resolve the additional magnetic field growth through the MRI, and conclude similarly that radiation still dominates the dynamics of the system with a radiation-to-magnetic pressure ratio of $\sim10^2$.

\subsection{Previous (R)HD simulations and other models}
\label{ss:otherworks}
Our simulations display several common features with previous (radiation) hydrodynamic works across a varied parameter space. We find that the interactions of the stream with itself and with the accumulating gas dominate the radiative emission, with luminosities reaching slightly super-Eddington values ($L/L_{Edd}\sim1-1.3$) of $\sim4-6\times10^{44}$ erg s$^{-1}$ around 7 days later. This prompt rise may be due to our constant fallback rate, which is meant to probe the TDE luminosity peak, rather than the full light curve. These luminosities are in agreement with previous radiation hydrodynamic simulations \citep{Jiang2016,Huang2023,Huang2024,Steinberg2024} and with observed TDE luminosities \citep{Gezari2021}. These stream-stream and stream-disk interactions launch fast radiatively-driven outflows and inefficiently circularize the gas. We find that the accretion flow remains mostly eccentric $e\sim0.4-0.5$ consistent with the $e\sim0.3-0.5$ range found in other works \citep{Shiokawa2015,Ryu2020,Steinberg2024,Huang2024}.

We find that the accretion flow has a polar-angle dependent density structure (see Figure \ref{fig:12}), reminiscent of that found in accretion-powered models \citep{Dai2018, Thomsen2022}. However, at this stage, we find that the accretion component of the radiative emission is subdominant to that produced by shocks. In addition, we find an azimuthal dependence to the density structure, established by the collision-driven outflows and subsequent inflows. This will result in larger optical depths further from the midplane for lines of sight in the vicinity of the collision zone, compared to those that would be observed through the opposite line of sight, near the pericenter. Our simulations suggest that most viewing angles will detect optical/UV emission, as the outflows remain optically thick to atomic transitions out to large radii. We find that the forming accretion disk remains largely eccentric and may remain so long after the collisions cease to be dynamically important \citep{Ryu2023}. This contrasts with the promptly formed super-Eddington Keplerian accretion disk model, in which larger radiative efficiencies may be expected \citep{Chan2024}. Importantly, in our resolved RMHD simulations, we find no evidence for significant viscous dissipation through the MRI that would accelerate circularization in the first seven days after the initial stream-stream collision. 

 Due to computational costs, we had to forgo the actual disruption of the star and, therefore, our initial conditions adopt simplifying assumptions. In particular, the time-dependence of the fallback rate should be taken into account as this may impact the dissipation in stream-stream and stream-disk collisions and the resulting outflows which reprocess radiation. This is explored in a companion work (Huang et al. in prep).
 
 Additionally, we approximate the general relativistic effects of a Schwarzschild spacetime using a generalized Newtonian potential. For spinning black holes, nodal precession may result in delayed circularization \citep{Guillochon2015}. Moreover, in the presence of a large magnetic flux around a spinning black hole, a jet may be powered through the Blandford$-$Znajek process \citep{Blandford1977,Tchekhovskoy2011}. Such a jet may alter the temperature structure of the system. The large magnetic flux required may be accumulated through the magnetic flux mixing of the debris stream with that of a fossil disk through hydrodynamic instabilities \citep{Tchekhovskoy2014,Kelley2014}. \cite{Curd2023} found in their GRMHD simulations that a magnetically arrested (MAD) state may be sustained and jets launched when a TDE debris stream interacts with a pre-existing MAD disk. However, GRRMHD simulations of TDEs in isolation show that when a weak field is present, no jet is launched \citep{Curd2021}.

 Lastly, our choice of magnetic field strength was motivated by computational constraints rather than by what may be expected of the evolution of a TDE of a Sun-like star with an average $\sim1$ G field \citep{Bonnerot2017_2}. A large magnetic field was necessary to ensure the resolution of the MRI, which has been suggested to possibly aid circularization in TDE systems \citep{Guillochon2014,Andalman2022, Tamilan2024}. We find that this effect is not observed in our simulations, since dissipation is dominated by collisions and angular momentum redistribution involves complex dynamics between collision-driven outflows and inflows (see Figure \ref{fig:a1}). We also do not observe the strengthening of collision shocks from angular momentum loss by magnetic stresses \citep{Bonnerot2017}. Additionally, although we inject a strong magnetic field, we do not find magnetically-driven winds \citep{Tamilan2024} in our simulations, since radiation still dominates the dynamics.

\section{Conclusions}
\label{s:conclusions}
We ran three radiation magnetohydrodynamic simulations using \texttt{Athena++} to explore the role of magnetic fields on the evolution of the fallback stellar debris after a tidal disruption event. These consisted of modeling the fallback stream as an injection of gas at a fixed rate of $\dot{M}_{fb}=10\,\dot{M}_{Edd}$ for the disruption ($\beta=1.73$) of a Sun-like star around $M_{BH}=3\times10^6\,M_{\odot}$ black hole. Two simulations included magnetized streams with the same average initial field strength $\sim2600$ G, but different stream magnetic field topologies, and one evolved a debris stream with zero magnetic fields. We found that the main effect of large magnetic fields is to thicken the stream as it falls back toward the black hole due to the large magnetic pressure support. Therefore, for the magnetized systems the stream collision zone was less dense compared to that for the nonmagnetized stream, which resulted in a less efficient acceleration of the gas by the radiation produced during the initial stream-stream collisions. Consequently, at very early times higher accretion rates and smaller mass outflow rates were measured in the MHD runs. As gas accumulated near the black hole, the stream began interacting instead with the forming accretion flow. Subsequently, the mass accretion and mass outflow rates measured varied within very similar ranges, with values $\sim30-50\%$ and $\sim7-12\%$ of the fallback rate, respectively, 7 days after the initial stream-stream collisions. The energy accreted through the inner boundary is largely kinetic, while the outgoing energy is dominated by radiation. By the end of the runs, the accreted radiative energies, measured at $r=3\,r_s$ reach $\sim 1-1.5 \,L_{Edd}$ and the outgoing luminosities range between $L\sim1-1.3 \,L_{Edd}$. The accreted and outflowing magnetic and thermal energies constitute only a small percentage of the energy content of the flows.

We find that outflows dominate the early evolution after the stream-stream collision, in agreement with previous work \citep{Huang2024}. These radiation-driven outflows have speeds that vary between $\sim0.05-0.18\,c$ for the unbound gas and between $\sim0.005-0.05\,c$ for the bound gas. We find an azimuthal dependence to the distribution of the outflowing gas that arises because the driving of outflows is strongest on the side of the flow where the collisions primarily occur. The transition to the accretion-dominated stage, in which $\dot{M}_{acc}>\dot{M}_{out}$, happens around $\sim3$ days after the initial stream-stream collision. However, the time when this transition happens will depend on the rate of dissipation, which is determined by the stream structure and the time-dependence of the fallback rate. 

We measure values for the normalized Reynolds and Maxwell stresses of $\alpha_{Re}\sim0.02-0.3$ and $\alpha_{M}\sim0.002-0.006$, respectively. However, we show that at this stage in the evolution, vertical and radial angular momentum fluxes induced by the persistent outflows and inflows dominate the angular momentum redistribution, therefore, regulating accretion. Given that this collision-driven dynamics dominates the flow, we do not observe a significant difference in the accretion rate between magnetized and nonmagnetized systems. Although we resolve the MRI, the angular momentum transport through this channel is subdominant to the dissipation and angular momentum redistribution due to the stream-stream and stream-disk collision shocks. We find, in agreement with previous work, that this dominant dissipation mechanism powers the TDE luminosity, while leaving an eccentric accretion flow, $e\sim0.2-0.3$ for $r\lesssim8r_s$ and $e\sim0.4-0.5$ for $10\lesssim r\,(r_s)\lesssim 50$. These collisions rapidly create a radiation-dominated system, with radiation pressures $\sim2$ and $\sim3-4$ orders of magnitude larger than the magnetic and thermal pressures, respectively. The resultant magnetic field structure in the accretion flow is preferentially toroidal with numerous reversals.

We find that the viewing-angle effect often invoked to explain TDE emission from super-Eddington accretion disks \citep{Dai2018}, may also exist in collision-powered models. The dependence of the density structure with distance from the midplane in our models is compatible with that required for the viewing-angle effect, which relies in diffuse outflows observed near the poles, while dense flows are observed near the midplane. In addition, we find an azimuthal dependence to the density structure, where larger scale heights are observed for lines of sight near the collision zone and smaller ones for those pointing at the pericenter, consistent with previous work \citep{Curd2021,Huang2024}. Throughout azimuth and radius, the vertical structure is supported by radiation. At this stage, the outflows remain optically thick for most viewing angles, and radiative emission is expected in the optical/UV bands. X-ray emission may be observable for some lines of sight as the outflows subside. Radiative emission across different wavebands for a longer time evolution is explored in a companion work (Huang et al. in prep).

In conclusion, in the first seven days after the initial stream-stream collision shock, we find no evidence in our resolved simulations for enhanced dissipation through the MRI, strengthened collision shocks due to magnetic stresses, nor magnetically-driven winds. As the collisions become less dynamically important, an eccentric accretion disk may result, in which the MRI may redistribute angular momentum more significantly and modify the eccentricity distribution \citep{Chan2022, Chan2024}. Exploring this would require following the evolution for possibly several weeks, which becomes computationally prohibitive and increasingly sensitive to our constant fallback rate assumption. Moreover, turbulence may still be sustained by returning bound outflows and keep regulating accretion instead \citep{Sadowski2016}. Given that dissipation through collision shocks dictates the formation and evolution of the accretion flow following a TDE, future studies should include more precise modeling of the time-dependent debris stream structure. In future work, we will focus on modeling the disruption of the star, as well as the debris evolution, and follow the accretion flow formation in general relativity.

\begin{acknowledgements}
    We thank the anonymous referee for providing helpful comments that improved this work. We thank Kengo Tomida for his helpful collaboration in solving initial problems with the simulation setup. We also thank Phil Chang, Edwin (Chi-Ho) Chan, Eric Coughlin, and Lizhong Zhang for useful discussions. We thank Zhaohuan Zhu for providing the opacity tables. XH is supported by the Sherman Fairchild Postdoctoral Fellowship at the California Institute of Technology. This work used Stampede 3 at Texas Advanced Computing Center through allocation TG-PHY240041 from the Advanced Cyber infrastructure Coordination Ecosystem: Services $\&$ Support (ACCESS) program. This work also made use of UVA's High-Performance Computing systems Rivanna and Afton. Support for this work was provided by the National Science Foundation under grant 2307886. XH appreciates the hospitality and interactions during the tde24 workshop, which is supported by the NSF grant PHY-2309135 to the Kavli Institute for Theoretical Physics (KITP). The Center for Computational Astrophysics at the Flatiron Institute is supported by the Simons Foundation. We acknowledge funding from the Virginia Institute for Theoretical Astrophysics (VITA), supported by the College and Graduate School of Arts and Sciences at the University of Virginia.
\end{acknowledgements}

\bibliography{references}

\end{document}